\documentclass[pra,twocolumn,amsmath,amssymb,floatfix,reprint,footinbib,superscriptaddress,longbibliography,showkeys]{revtex4-2}
\usepackage{dcolumn}
\usepackage{graphicx}
\usepackage{mathrsfs}
\usepackage{mdwlist}
\usepackage{subfigure}
\usepackage{booktabs}
\usepackage{amsmath}
\usepackage{dsfont}
\usepackage{amstext}
\usepackage{amssymb}
\usepackage{amsbsy}
\usepackage{bbm}
\usepackage{amsthm}
\usepackage{graphicx}
\usepackage{textcomp}
\usepackage{multirow}
\usepackage{color}
\usepackage[colorlinks,citecolor=blue]{hyperref}
\definecolor{Dgreen}{RGB}{0, 100, 0}
\usepackage{url}
\usepackage[colorlinks]{hyperref}
\hypersetup{%
    plainpages=true,
    breaklinks=true,  
    hypertexnames=false,  
    pageanchor=true,
    colorlinks=true,
    linkcolor={blue},
    citecolor={red},
    urlcolor={blue},
    anchorcolor={black}
}

\begin{document}

\preprint{APS/123-QED}

\title{Nonadiabatic geometric quantum computation with cat qubits via invariant-based reverse engineering}

\author{Yi-Hao Kang }\thanks{The first two authors contributed equally.}
\affiliation{Department of Physics, Harbin Institute of Technology,
Harbin 150001, China}
\affiliation{Fujian Key Laboratory of Quantum
Information and Quantum Optics, Fuzhou University, Fuzhou 350116,
China}

\author{Ye-Hong Chen}\thanks{The first two authors contributed equally.}
\affiliation{Theoretical Quantum Physics Laboratory, RIKEN Cluster
for Pioneering Research, Wako-shi, Saitama 351-0198, Japan}

\author{Xin Wang}
\affiliation{Institute of Quantum Optics and Quantum Information, School of Science, Xi'an Jiaotong University, Xi'an 710049, China}


\author{Jie Song}\thanks{jsong@hit.edu.cn}
\affiliation{Department of Physics, Harbin Institute of Technology,
Harbin 150001, China}

\author{Yan Xia}\thanks{xia-208@163.com}
\affiliation{Fujian Key Laboratory of Quantum Information and Quantum Optics, Fuzhou University, Fuzhou 350116, China}
\affiliation{Department of Physics, Fuzhou University, Fuzhou 350116, China}

\author{Adam Miranowicz}
\affiliation{Theoretical Quantum Physics Laboratory, RIKEN Cluster for Pioneering Research, Wako-shi, Saitama 351-0198, Japan}
\affiliation{Institute of Spintronics and Quantum Information,
    Faculty of Physics, Adam Mickiewicz University, 61-614 Pozna\'n, Poland}

\author{Shi-Biao Zheng}
\affiliation{Fujian Key Laboratory of Quantum Information and Quantum Optics, Fuzhou University, Fuzhou 350116, China}
\affiliation{Department of Physics, Fuzhou University, Fuzhou 350116, China}

\author{Franco Nori}
\affiliation{Theoretical Quantum Physics Laboratory, RIKEN Cluster
for Pioneering Research, Wako-shi, Saitama 351-0198, Japan}
\affiliation{RIKEN Center for Quantum Computing (RQC), Wako-shi, Saitama 351-0198, Japan}
\affiliation{Department of Physics, University of Michigan, Ann
Arbor, Michigan 48109-1040, USA}

\date{\today}

\begin{abstract}
We propose a protocol to realize nonadiabatic geometric quantum
computation of small-amplitude Schr{\"o}dinger cat qubits via
invariant-based reverse engineering. We consider a system with a
two-photon driven Kerr nonlinearity, which can generate a pair of
dressed even and odd coherent states (i.e., Schr{\"o}dinger cat
states) for fault-tolerant quantum computations. An additional
coherent field is applied to linearly drive a cavity mode, to induce
oscillations between dressed cat states. By designing this linear
drive with invariant-based reverse engineering, we show how to implement nonadiabatic
geometric quantum computation with cat qubits.
The performance of the protocol is estimated by taking into account
the influence of systematic errors, additive white Gaussian noise, $1/f$ noise,
and decoherence including photon loss and dephasing. Numerical
results demonstrate that our protocol is robust against these
negative factors. Therefore, this protocol may provide a feasible
method for nonadiabatic geometric quantum computation in bosonic
systems.
\end{abstract}

\maketitle


\section{Introduction}

Quantum coherence and quantum entanglement are arguably the most
fascinating properties of quantum mechanics
\cite{NielsenBook,BennettNat404,GribbinBook,Zhong2020}.
These are the main resources for quantum information processing~\cite{NielsenBook,TaylorBook} and quantum technologies of second
generation~\cite{HarocheBook}. Their recent applications include:
demonstrations of quantum advantage using superconducting
programmable processors~\cite{Arute2019} or boson sampling with
squeezed states of photons \cite{Zhong2020}, a quantum
communication network over 4,600~km~\cite{Chen2021}, and
quantum-enhanced gravitational-wave detectors using squeezed
vacuum~\cite{Aasi2013,Grote2013,Acernese2019,Tse2019}.
As a very important subfield of quantum information processing, quantum computation has shown a potentially great power in solving many specific problems
\cite{EkertRMP68,GroverPRL79}. In a practical
implementation of quantum computation, quantum algorithms are
usually (but not always, e.g., in quantum annealing) designed as a sequence of quantum gates. Therefore,
high-fidelity quantum gates are essential elements of quantum
computation. Unfortunately, experimental imperfections, including
operational errors, parameter fluctuations, and environment-induced
decoherence, may affect the desired dynamics, limiting the fidelities
of quantum gates. The problem how to overcome these experimental imperfections
has to be solved for the constructions of practical quantum computers.

Because geometric
phases are determined by the global geometric properties of the
evolution paths,
geometric quantum computation \cite{Cen2006,ZSLPRA72} has shown
robustness against local parameter fluctuations over a cyclic
evolution \cite{ZSLPRA72,SjoqvistPhys1,LQXPRA101}. As an extension of geometric quantum computation,
holonomic quantum computation
\cite{ZanardiPLA264,DLMSci292,WLAPRL95} based on non-Abelian
geometric phases can be used to construct a universal set of
single-qubit gates and several two-qubit entangling gates. Early implementations of geometric
quantum computation involve adiabatic evolutions to suppress
transitions between different eigenvectors of the Hamiltonian. This
makes the evolution slow, and decoherence may destroy these geometric gates \cite{Ashhab2006,Ashhab2010,Zheng2010,Wilson2012,Zhang2017}.

To speed up the evolution,
nonadiabatic geometric quantum computation (NGQC)
\cite{ZSLPRL89,SjoqvistNJP14,XGFPRL109} was proposed. Note that NGQC is also enabled by geometric
phases, thus inheriting robustness against local parameter
fluctuations. Moreover, compared with adiabatic geometric quantum
computation \cite{ZanardiPLA264,DLMSci292,WLAPRL95}, NGQC is faster
because the evolution is beyond the adiabatic limit. In addition,
NGQC is compatible with various quantum optimized control techniques, such as reverse engineering
\cite{GaraotPRA89,LYCPRA97,OdelinRMP91,KYHAQT3} and
single-shot-shaped pulses \cite{Wei2008,DaemsPRL111,DammePRA96}.
Such techniques provide flexible ways in designing
evolution paths for NGQC, reducing the number of auxiliary
levels and sensitivity to certain types of control errors
\cite{DYXQUTE2}. Because of the above advantages of NGQC, in the past decades, robust
quantum computation has been discussed in theory
\cite{XZYPRA94,XGFPRA95,KYHPRA102} and successfully demonstrated in
experiments \cite{AbdumalikovNat496,ZBBPRL119,XYPRL121}.

Recent research has shown \cite{HuLNP15,MaYNP16,ZhangPRA2018} that
encoding quantum information in logical qubits is promising to
protect quantum computation from errors. For the realization of logical
qubits, bosonic systems are promising candidates, which can be
constructed by quantized fields in, e.g., resonators, mechanical
oscillators, and superconducting Josephson junctions
\cite{MirrahimiNJP16,GiovanniniSci347,Purinpj3,GuillaudPRX9,GrimmNat584}.
The Schr{\"o}dinger cat states of bosons
\cite{Dodonov1974} have shown applications in
quantum computation in the early
2000s \cite{Ralph2003,Gilchrist2004}. Subsequently, it has been shown
that the Schr{\"o}dinger cat states
\cite{Liu2005,Kira2011,Gribbin2013,Puri2019,CYHprl2021,Qin2021} can be used to
construct types of useful error-correction codes
\cite{GottesmanPRA2001,TerhalRMP87,Gaitan2008,Daniel2009,Mirrahimi2016,AlbertPRL116,MichaelPRX2016,AlbertPRA2018,AlbertQST4,Litinski2019,GrimsmoPRX2020,CYHPRR2021,Chamberland2020,Ma2021,CaiFR2021},
providing protection against cavity dephasing
\cite{TerhalRMP87,AlbertPRL116}, and thus have attracted much interest.

Recently, based on cat codes, various protocols
\cite{GuillaudPRX9,GrimmNat584,chamberland2020building,Purinpj3,MirrahimiNJP16,GiovanniniSci347,CYHprl2021}
for preparing, stabilizing, and manipulating cat qubits have
been put forward. Moreover, quantum computation \cite{GuillaudPRX9}
and adiabatic geometric quantum control \cite{AlbertPRL116} of cat
qubits have also been considered. However, so far, only a few protocols \cite{AlbertPRL116,CYHarxiv2021} have been proposed to implement geometric computation
using cat qubits. Because of the difficulty to arbitrarily manipulate a bosonic mode, it is still a challenge to realize NGQC using
bosonic cat qubits, which are both robust and fault tolerant.

In this manuscript, we propose to use cat qubits to implement NGQC
via invariant-based reverse engineering.
To construct cat qubits, a two-photon driven Kerr nonlinearity is
used to restrict the evolution of cavity modes to a subspace spanned
by a pair of cat states. We apply an additional coherent field to
linearly drive a cavity mode in order to induce oscillations between
dressed cat states. With the control fields designed by
invariant-based reverse engineering, the system can have a cycling
evolution, which acquire only pure geometric phases. Hence, NGQC
with cat qubits can be implemented.

An amplitude-amplification
method, using light squeezing \cite{SekatskiPRL103,Qin2019}, is
applied to increase the distinguishability of different cat states,
so that it can be easier to detect input and output states in
practice. Moreover, two-qubit quantum gates of cat qubits are also
considered by using couplings between two cavity modes. Controlled
two-qubit geometric quantum gates can be implemented almost
perfectly.

Finally, the performance of the protocol in the presence
of systematic errors, additive white Gaussian noise (AWGN), $1/f$ noise, and decoherence
(including photon loss and dephasing) are investigated via numerical
simulations. Our results indicate that the protocol is robust
against these negative factors.

The article is organized as follows. In Sec. II, we briefly
introduce the basic theory for invariant-based NGQC. In Sec. III, we
describe how to implement single- and two-qubit NGQC with cat
qubits. In Sec. IV, we consider experimental imperfections and
estimate the performance of the protocol via numerical simulations.
In Sec. V, we introduce an amplification method based on quadrature
squeezing for the amplitudes of cat states, so that the detection of
input and output states can be performed easily. In Sec. VI, we
discuss a possible implementation of our protocol using a
superconducting quantum parametron. Finally, our conclusions are
given in Sec. VII. Appendix A includes a derivation of a dynamic
invariant and the choice of parameters for eliminating dynamical
phases.

\section{\label{II}Nonadiabatic geometric quantum computation based on a dynamic invariant}

For details, we first recall the Lewis-Riesenfeld invariant theory
\cite{LewisJMP10}. Assuming that a physical system is described by a
Hamiltonian $H(t)$, a Hermitian operator $I(t)$ satisfies the
following equation $(\hbar=1)$
\begin{equation}\label{lr1}
i\frac{\partial}{\partial t}I(t)-[H(t),I(t)]=0.
\end{equation}
For a non-degenerate eigenvector $|\phi_l(t)\rangle$ of $I(t)$,
$|\psi_l(t)\rangle=\exp[i\alpha_l(t)]\,|\phi_l(t)\rangle$ is a
solution of the time-dependent Schr\"{o}dinger equation
$i|\dot{\psi}(t)\rangle=H(t)|\psi(t)\rangle$. Here, $\alpha_l(t)$ is
the Lewis-Riesenfeld phase defined as
\begin{equation}\label{lr2}
\alpha_l(t)=\int_{0}^{t}\langle\phi_l(\tau)|\left[i\frac{\partial}{\partial
    \tau}-H(\tau)\right]|\phi_l(\tau)\rangle d\tau.
\end{equation}

To realize NGQC, one can select a set of time-dependent vectors,
$\{|\phi_l(t)\rangle\}$, spanning a computational subspace
$\mathcal{S}$. According to Ref.~\cite{LBJPRL123},
$\{|\phi_l(t)\rangle\}$ should satisfy the three conditions: (i) the
cyclic evolution condition $|\phi_l(0)\rangle=|\phi_l(T)\rangle$
with $T$ being the total operation time; (ii) the von Neumann
equation
\begin{equation}\label{lradd1}
\dot{\Xi}_l(t)=-i[H(t),\Xi_l(t)],
\end{equation}
with $\Xi_l(t)=|\phi_l(t)\rangle\langle\phi_l(t)|$; and (iii) annihilation
of the dynamical phase
\begin{equation}\label{lradd2}
\vartheta_l(T)=-\int_0^{T}\langle\phi_l(t)|H(t)|\phi_l(t)\rangle\,
dt=0.
\end{equation}
When satisfying the three conditions, the evolution in
subspace $\mathcal{S}$ can be described by
\begin{equation}\label{lr3}
U(T,0)=\sum\limits_l\exp[i\Theta_l(T)]\,\Xi(0),
\end{equation}
with a pure geometric phase
\begin{equation}\label{lr4}
\Theta_l(T)=\int_0^{T}\langle\phi_l(t)|i\frac{\partial}{\partial
    t}|\phi_{l}(t)\rangle\,dt.
\end{equation}
Ref.~\cite{KYHPRA101} has shown that, a non-degenerate eigenvector
$|\phi_l(t)\rangle$ of an invariant obeys the von Neumann equation given
in Eq.~(\ref{lradd1}). Therefore, when the parameters of $I(t)$ are
designed with the cycling boundary conditions and the dynamical part of
the Lewis-Riesenfeld phase is eliminated, the three
conditions are satisfied. In this case, one can implement {NGQC} with a
dynamic invariant $I(t)$.

\section{Nonadiabatic geometric quantum computation of cat qubits}\label{III}

\subsection{Arbitrary single-qubit gates}

\subsubsection{Hamiltonian and evolution operator of a single resonator}

We consider a resonant single-mode two-photon (i.e., quadrature) squeezing drive applied to a Kerr-nonlinear resonator. In the frame rotating at the resonator
frequency $\omega_1$, the system is described by the Hamiltonian
\cite{GrimmNat584,Purinpj3}
\begin{equation}\label{e1}
H_{\mathrm{cat}}=-Ka_{1}^{\dag2}a_{1}^2+\epsilon_2\left(e^{2i\xi}a_{1}^{\dag2}+e^{-2i\xi}a_{1}^2\right),
\end{equation}
where $K$ is the Kerr nonlinearity, $a_{1}$ ($a_1^\dag$) is the annihilation (creation) operator of the resonator (cavity) mode, $\epsilon_2$ is the
strength of the two-photon drive assumed here to be real, and $\xi$ is its phase. The coherent states
$|\pm\alpha\rangle_{1}$ (where $\alpha=\sqrt{\epsilon_2/K} e^{i\xi}$ is the complex amplitude) are two
degenerate eigenstates of $H_{\mathrm{cat}}$.
Therefore, the even ($|\mathcal{C}_+\rangle_{1}$) and odd ($|\mathcal{C}_-\rangle_{1}$) coherent states, often referred to as  Schr{\"o}dinger cat states, which are defined as
\begin{align}\label{e2}
|\mathcal{C}_\pm\rangle_{1}=\frac{1}{\sqrt{\mathcal{N}_\pm}}\left(|\alpha\rangle_{1}\pm|-\alpha\rangle_{1}\right),
\end{align}
are two orthonormal degenerate eigenstates of $H_{\mathrm{cat}}$,
where $\mathcal{N}_\pm=2[1\pm\exp(-2|\alpha|^2)]$ are the normalized
coefficients. The total Hamiltonian
\begin{equation}\label{e2.5}
H_{\rm{tot}}(t)=H_{\mathrm{cat}}+H_c,
\end{equation}
includes a control Hamiltonian defined as
\cite{GrimmNat584}
\begin{equation}\label{e3}
H_{c}(t)=\chi(t)a_{1}^\dag
a_{1}+\epsilon(t)a_{1}^\dag+\epsilon^*(t)a_{1},
\end{equation}
with $\chi(t)$ and $\epsilon(t)$ being the detuning and
strength of a single-photon drive, respectively.

When the energy gap $E_{\mathrm{gap}}$
between the cat states $|\mathcal{C}_\pm\rangle_{1}$ and their nearest
eigenstate of $H_{\mathrm{cat}}$ is much larger than $\chi(t)$ and
$\epsilon(t)$, the system can be restricted to the subspace
$\mathcal{S}$ spanned by $|\mathcal{C}_\pm\rangle_{1}$. We can
accordingly use the cat states to define the Pauli matrices as
\begin{align}
  \sigma_{x}&=\sigma_{+}+\sigma_{-},\ \ \ \ \ \ \ \ \
  \sigma_{y}=i(\sigma_{-}-\sigma_{+}), \cr
  \sigma_{z}&=\sigma_{+}\sigma_{-}-\sigma_{-}\sigma_{+},\ \ \
  \vec{\sigma}=(\sigma_{x},\sigma_{y},\sigma_{z}),
\end{align}
in terms of the raising ($\sigma_{+}$) and lowering ($\sigma_{-}$) qubit operators,
\begin{equation}\label{e3.5}
\sigma_{+}=|\mathcal{C}_{+}\rangle_{1}\langle \mathcal{C}_{-}|,\ \ \ \ \sigma_{-}=|\mathcal{C}_{-}\rangle_{1}\langle \mathcal{C}_{+}|
\end{equation}
Then, the
Hamiltonian of the system can be simplified to
$H_c(t)=\vec{\Omega}(t)\cdot\vec{\sigma}$, where, $\vec{\Omega}(t)=[\Omega_{x}(t),\Omega_{y}(t),\Omega_{z}(t)]$
is a set of driving amplitudes to be determined.
{We find that $I(t)=\vec{\zeta}(t)\cdot\vec{\sigma}$ is a dynamic invariant, where $\vec{\zeta}(t)=[\zeta_x(t),\zeta_y(t),\zeta_z(t)]$ is the three-dimensional time-dependent vector satisfying the conditions
$\dot{\vec{\zeta}}(t)=2\vec{\Omega}(t)\times\vec{\zeta}(t)$ and
$|\vec{\zeta}(t)|=\mathrm{const}$ (see Appendix A for details). The three components $[\zeta_x(t),\zeta_y(t),\zeta_z(t)]$ of $\vec{\zeta}(t)$ denote the projections of the dynamic invariant $I(t)$ along the directions $(\sigma_{x},\sigma_{y},\sigma_{z})$ in the SU(2) algebra.}

\begin{figure}[b]
\centering\scalebox{0.5}{\includegraphics{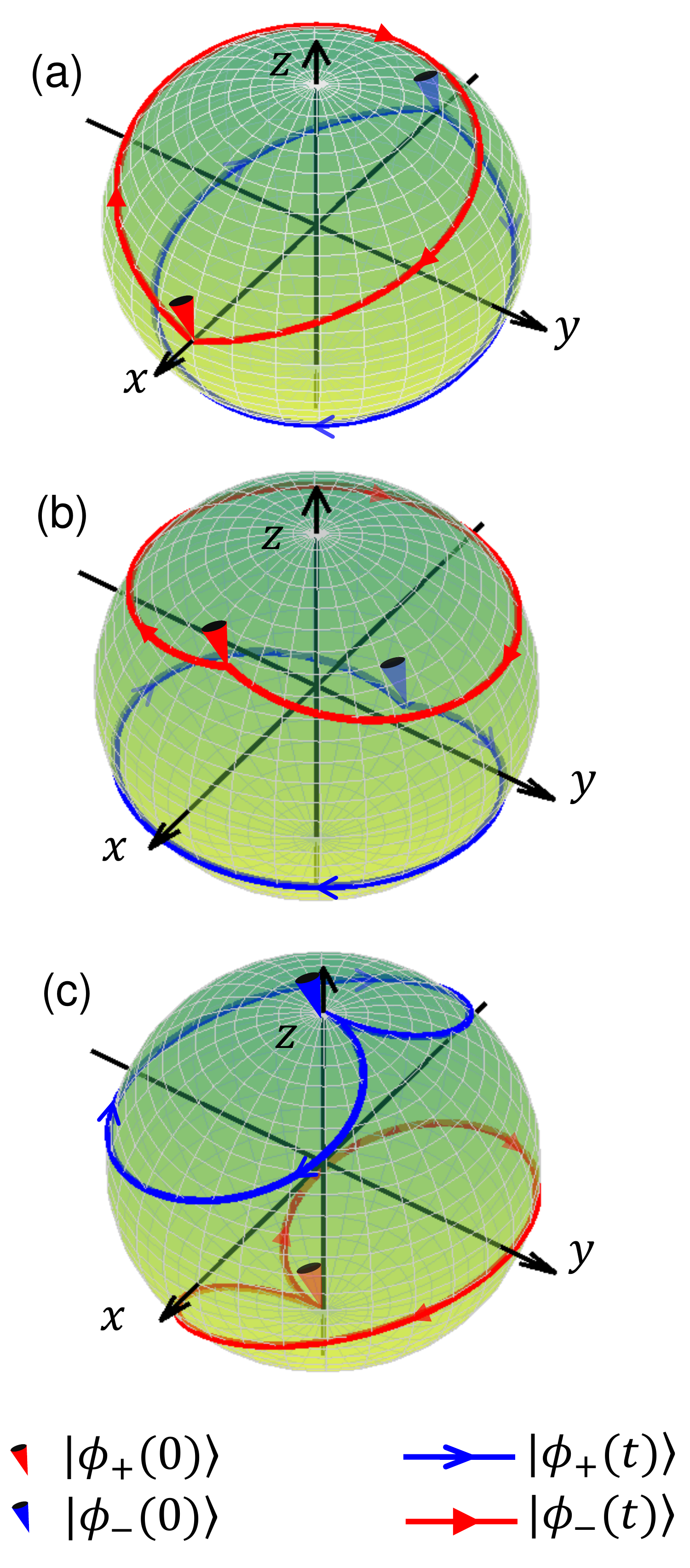}}
\caption{\label{fig1} Trajectories of the eigenvectors $|\phi_+(t)\rangle$ (red-solid curve) and $|\phi_-(t)\rangle$, defined in Eq.~(\ref{eq1-5}), on the Bloch sphere in the implementations of
        (a) the NOT gate, (b) the Hadamard gate, and (c) the $\pi$-phase
        gate. Parameters are listed in Table~\ref{tab1}.}
\end{figure}

By introducing two time-dependent dimensionless parameters $\eta$ and
$\mu$, we can parametrize $\vec{\zeta}(t)$ as
$(\sin{\eta}\sin{\mu},\cos{\eta}\sin{\mu},\cos{\mu})$
and the eigenvectors of the dynamic invariant $I(t)$ can be derived
as
\begin{align}\label{eq1-5}
&|\phi_{+}(t)\rangle=\cos\frac{\mu}{2}|\mathcal{C}_+\rangle_{1}+i\exp(-i\eta)\sin\frac{\mu}{2}|\mathcal{C}_-\rangle_{1},\cr
&|\phi_{-}(t)\rangle=i\exp(i\eta)\sin\frac{\mu}{2}|\mathcal{C}_+\rangle_{1}+\cos\frac{\mu}{2}|\mathcal{C}_-\rangle_{1}.
\end{align}
We can design the parameters $\epsilon(t)$ and $\chi(t)$ as
\begin{align}\label{eq1-4}
\mathrm{Re}[\epsilon(t),\xi]=&~\frac{\sqrt{\mathcal{N}_+\mathcal{N}_-}}{8\alpha}
\left(\Omega_{x}\cos\xi+e^{2|\alpha|^2}\Omega_{y}\sin\xi\right), \cr
\mathrm{Im}[\epsilon(t),\xi]=&~\mathrm{Re}[\epsilon(t),\xi-\pi/2], \cr
   \chi(t)=&~\frac{(\dot{\eta}\sin^{2}\mu )\mathcal{N}_{+}\mathcal{N}_{-}}{[(\mathcal{N}_{+}^{2}-\mathcal{N}_{-}^{2})|\alpha|^{2}]},
\end{align}
with effective driving amplitudes (see Appendix A for details)
\begin{align}
  \Omega_{x}=&~\frac{1}{2}\left[{\dot{\eta}}\sin\eta\sin(2\mu)-2\dot{\mu}\cos\eta\right], \cr
  \Omega_{y}=&~\frac{1}{2}\left[{\dot{\eta}}\cos\eta\sin(2\mu)+2\dot{\mu}\sin\eta\right].
\end{align}
Both dynamical phases, acquired by the
eigenvectors shown in Eq.~(\ref{eq1-5}), vanish due to $\langle
\phi_{\pm}(t)|H_{c}(t)|\phi_{\pm}(t)\rangle=0$, while the
geometric phases acquired by $|\phi_\pm(t)\rangle$, defined in Eq.~(\ref{eq1-5}), are
\begin{align}\label{eq1-6}
  \Theta_\pm(t)&\equiv~\int_0^{t}\langle\phi_{\pm}(\tau)|i\frac{\partial}{\partial
    \tau}|\phi_{\pm}(\tau)\rangle
    d\tau\cr&=~\pm\int_{0}^{t}\dot{\eta}\sin^2\left(\frac{\mu}{2}\right)d\tau.
\end{align}
According to Eq.~(\ref{lr3}), the evolution of the system in the
subspace $\mathcal{S}$, after a cycling evolution with period $T$, is
calculated as
\begin{align}\label{eq1-7}
&U_{s}(T,0)=\exp\left[i\theta\vec{\zeta}(0)\cdot\vec{\sigma}\right]\cr=&\left[%
\begin{array}{cccc}
\cos\theta+i\cos\mu_{0}\sin\theta & \exp(i\eta_{0})\sin\mu_{0}\sin\theta \\
-\exp(-i\eta_{0})\sin\mu_{0}\sin\theta & \cos\theta-i\cos\mu_{0}\sin\theta \\
\end{array}\right],\ \ \ %
\end{align}
where $\theta=\int_{0}^{T}\dot{\eta}\sin^2(\mu/2)\,d t$ is the
final geometric phase of the cycling evolution, and $\mu_0$
($\eta_0$) is the initial value of $\mu$ ($\eta$). The evolution
operator $U_{s}(T,0)$ represents a rotation on the Bloch sphere that can generate
arbitrary single-qubit gates \cite{LZTPRA93,JLNPRA100}. For a
cycling evolution, the parameters can be interpolated by
trigonometric functions as
\begin{align}\label{e14}
\mu=&~\mu_{0}+\Lambda\sin^2\left(\frac{\pi t}{T}\right),\cr
\eta=&~\eta_{0}+\pi\left[1-\cos\left(\frac{\pi t}{T}\right)\right],
\end{align}
where $\Lambda$ is an auxiliary parameter to be determined according to the requirements of
different gates.

\subsubsection{Examples of single-qubit-gate implementations}

\begin{figure}
\centering\scalebox{0.5}{\includegraphics{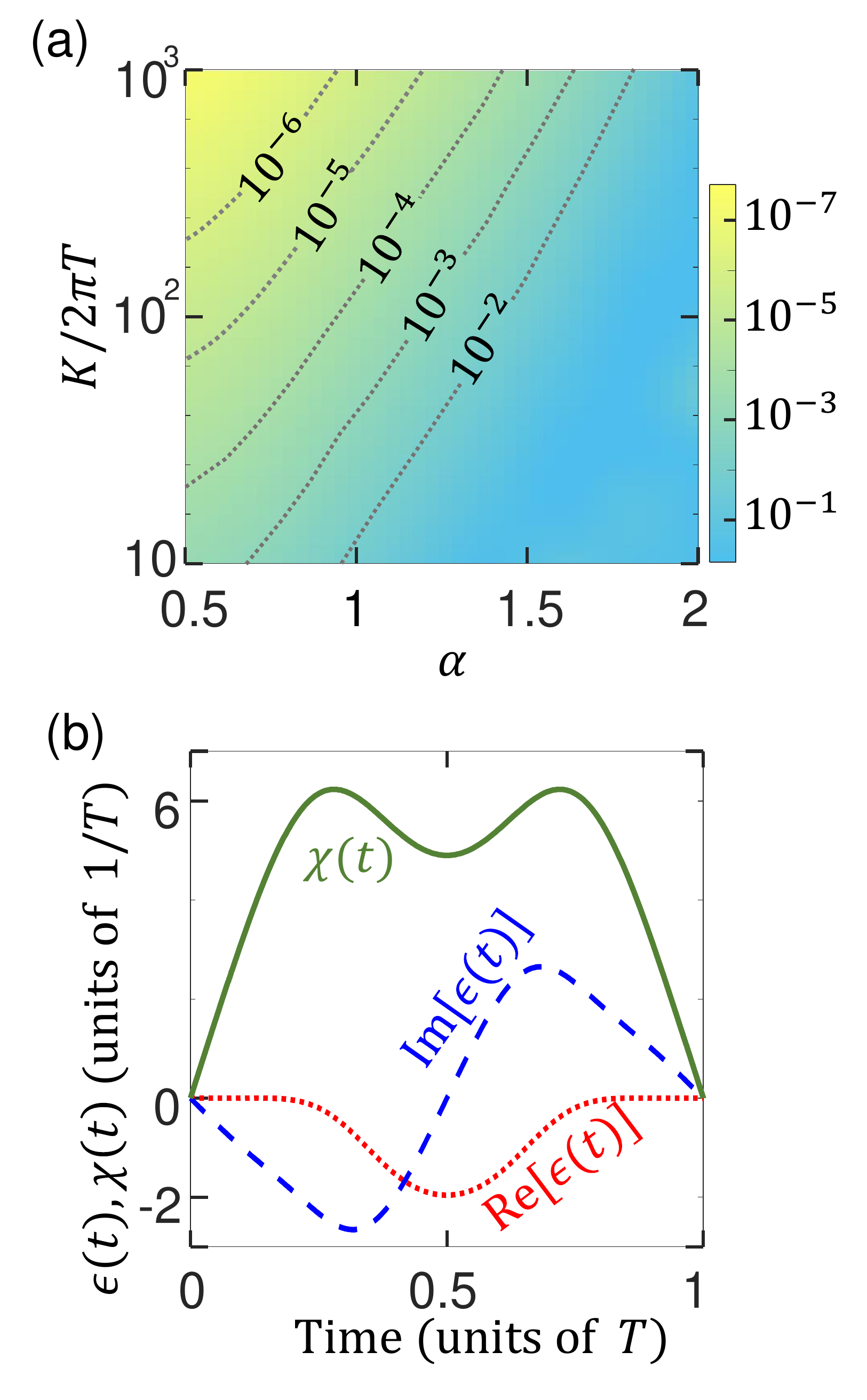}} \caption{(a) Infidelity $(1-\bar{F}_{\rm{NOT}})$
versus the amplitude $\alpha$ and the Kerr nonlinearity $K$ for our implementation of the NOT gate. (b) Time variations of the parameters $\chi(t)$, $\mathrm{Re}[\epsilon(t)]$,
and $\mathrm{Im}[\epsilon(t)]$, defined in Eq.~(\ref{eq1-4}). Parameters are listed in Table~\ref{tab1}.
}\label{fig2}
\end{figure}
\begin{figure*}
    \centering\scalebox{0.38}{\includegraphics{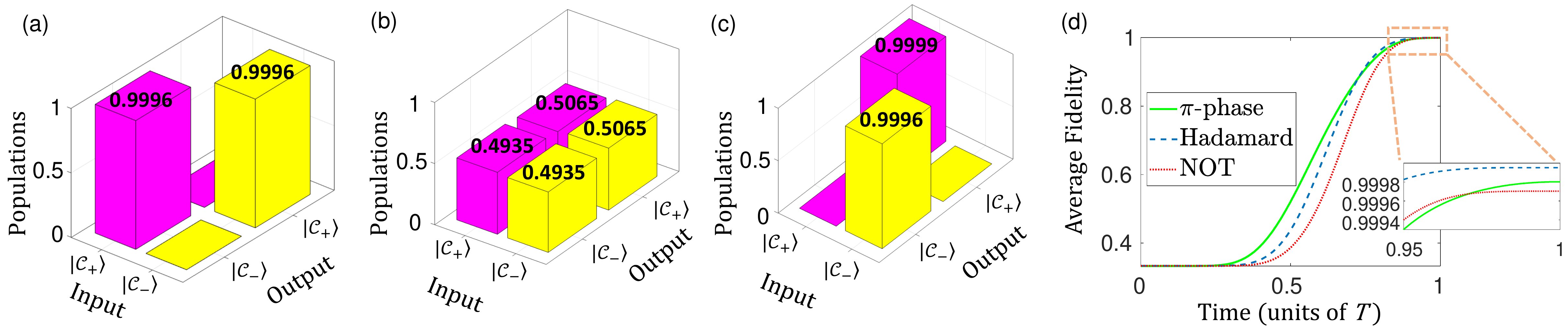}}
    \caption{Populations of different output states for different input
        states in the implementation of: (a) the NOT gate, (b) the Hadamard gate, and
        (c) the $\pi$-phase gate with parameters $K=2\pi\times12.5$~MHz and
        $|\alpha|=0.5$. {(d) Average fidelities of the NOT gate (red-dotted curve), the Hadamard gate (blue-dashed curve), and the $\pi$-phase gate (green-solid curve) with parameters $K=2\pi\times12.5$~MHz and
        $|\alpha|=0.5$.}}\label{fig3}
\end{figure*}

We now discuss how to use the evolution operator $U_{s}(T,0)$, given in
Eq.~(\ref{eq1-7}), to realize:

(i) the NOT gate, $U_{\rm{NOT}}=\sigma_{x}$;

(ii) the Hadamard gate, $U_{\mathrm{H}}=(\sigma_{z}+\sigma_{x})/\sqrt{2}$;

(iii) the arbitrary phase gate,
$U_{\rm{Phase}}(\theta)=\cos{\frac{\theta}{2}}\cdot\mathbbm{1}+i\sin{\frac{\theta}{2}}\cdot\sigma_{z}$,
where $\mathbbm{1}$ is the identity operator acting on the
cat qubit.

{To determine the initial values of $\mu_0$ and $\eta_0$, we can
exploit the evolution operator $U_s(T,0)$ shown in Eq.~(\ref{eq1-7}). Because
the system evolves through a cycling evolution [$|\phi_\pm(T)\rangle=|\phi_\pm(0)\rangle$], the evolution operator
only relies on the acquired geometric phase $\theta$ and the
initial values of $\mu_0$ and $\eta_0$. For example, to implement the
NOT gate, we should make the diagonal elements of the evolution operator $U_s(T,0)$ in Eq.~(\ref{eq1-7}) to become zeros. Therefore, we set $\theta=\mu_0=\pi/2$. In addition, the
off-diagonal elements of $U_s(T,0)$ should be equal to 1. Thus, we select
$\eta_0=\pi/2$. Dropping a global phase $\pi/2$, $U_s(T,0)$
becomes the NOT gate
\begin{equation}
U_{\mathrm{NOT}}=\left[%
\begin{array}{cc}
0 & 1 \\
1 & 0 \\
\end{array}\right].
\end{equation}
Moreover, to implement the
Hadamard gate, the matrix for $U_s(T,0)$ should be equal to
\begin{equation}
U_{\mathrm{H}}=\frac{1}{\sqrt{2}}\left[%
\begin{array}{cc}
1 & 1 \\
1 & -1 \\
\end{array}\right].
\end{equation}
Comparing $U_{\mathrm{H}}$ and $U_s(T,0)$ in Eq.~(\ref{eq1-7}), we find that
$\mu_0=\pi/4$ and $\eta_0=\theta=\pi/2$. In this case, up to a
global phase $\pi/2$, $U_s(T,0)$ becomes the
Hadamard gate.}

{Finally, for the $\pi$-phase gate, the off-diagonal
elements of $U_s(T,0)$ should vanish. Therefore, we set $\mu_0=0$.
In this case, $U_s(T,0)$ is independent of the value of $\eta_0$, so that
we can choose $\eta_0=0$ for simplicity. Omitting the global phase $\theta$, $U_s(T,0)$
becomes
\begin{equation}
U_{\theta}=\left[%
\begin{array}{cc}
1 & 1 \\
1 & e^{-2i\theta} \\
\end{array}\right].
\end{equation}
Then, the $\pi$-phase gate can be realized with $\theta=\pi/2$.
For the sake of clarity, the corresponding parameters to realize the three gates for $\theta=\pi/2$ are listed in Table~\ref{tab1}.}

{\renewcommand\arraystretch{1.2}
\begin{table}[b]
    \centering
    \caption{Parameters for our implementations of single-qubit gates}
    \label{tab1}
    \begin{tabular}{p{1.5cm}<{\centering}p{1.5cm}<{\centering}p{1.5cm}<{\centering}p{1.5cm}<{\centering}p{1.5cm}<{\centering}}
        \hline
        \hline
        gate & $\mu_{0}$ & $\eta_{0}$ & $\theta$ & $\Lambda$  \\
        \hline
        $U_{\rm{NOT}}$ & $\pi/2$ & $\pi/2$ & $\pi/2$  & $0.8089$ \\
        $U_{\mathrm{H}}$ & $\pi/4$ & $\pi/2$ & $\pi/2$  &  $0.3859$ \\
        $U_{\rm{Phase}}(\pi)$ & $0$ & $0$ & $\pi/2$  & $1.4669$ \\
        \hline
        \hline
    \end{tabular}
\end{table}
}

According to the parameters given in Table~\ref{tab1}, on the Bloch sphere in Fig.~\ref{fig1}, we plot the
trajectories of the eigenvectors $|\phi_{\pm}(t)\rangle$, i.e.,
\begin{align}
\vec{r}_{\pm}(t)=\sum_{k=x,y,z}\mathrm{Tr}[|\phi_{\pm}(t)\rangle\langle\phi_{\pm}(t)|\sigma_k]\vec{e}_k,
\end{align}
where
$\vec{e}_{k}$ is the unit direction vector along the $k$-axis. As
shown in each panel of Fig.~\ref{fig1}, both vectors
$|\phi_{\pm}(t)\rangle$ evolve along their cycling paths individually, and
the geometric phases acquired by them are equal to half of the solid
angles of the areas surrounded by the corresponding paths. In
addition, the solid angles of the paths of $\vec{r}_{\pm}(t)$ have opposite
signs, because the paths are on the upper and lower half spheres,
respectively. These numerical results are in agreement with the
theoretical results in Eq.~(\ref{eq1-6}).

The average fidelity of the gates over all possible initial states in
the subspace $\mathcal{S}$ can be calculated by
\cite{ZanardiPRA70,PedersenPLA367}
\begin{equation}\label{e150}
\bar{F}_{G}=\frac{1}{\mathcal{D}(\mathcal{D}+1)}\left[\mathrm{Tr}(MM^\dag)
+|\mathrm{Tr}(M)|^2\right],
\end{equation}
with
$M=\mathcal{P}_{c}U^\dag_{G}U_{1}\mathcal{P}_{c}$,
while $\mathcal{P}_{c}$ and $\mathcal{D}$ are the projector and
dimension of the computational subspace, respectively. The
subscript ``$G$'' denotes the desired gate, e.g.,
$G=\rm{NOT}$ when one wants to implement the NOT gate.
Figure~\ref{fig2}(a) shows the infidelity $(1-\bar{F}_{\rm{NOT}})$
versus the amplitude $\alpha$ and the Kerr nonlinearity $K$ for the NOT gate, as an example.

We find that the average fidelity $\bar{F}_{\rm{NOT}}$ decreases sharply when the
amplitude $\alpha$ increases. This effect can be understood because the
control parameters $\chi(t)$ and $\epsilon(t)$ increase with
$\exp(|\alpha|^{2})$ according to Eq.~(\ref{eq1-4}).
Consequently, the ratio between the energy gap and parameters of $H_{c}(t)$ reduces,
and the leakage to other eigenstates of $H_{\rm{cat}}$ becomes
significant. To manipulate the cat qubit with a larger $\alpha$, one
may increase the Kerr-nonlinearity $K$ and the strength
$\epsilon_2$ of the squeezing drive, but we should notice that $K$
and $\epsilon_2$ both have their upper limits in experiments
\cite{GrimmNat584}. We can also consider a longer
interaction time $T$ to reduce values of the parameters in
$H_{\mathrm{add}}(t)$, but such a long-time evolution may increase
the influence of decoherence.

For a realistic value of the Kerr nonlinearity
$K=2\pi\times12.5$~MHz \cite{GrimmNat584}
($E_{\mathrm{gap}}=161$~MHz), the parameters $\chi(t)$ and
$\epsilon(t)$ are shown in Fig.~\ref{fig2}(b), when the total
interaction time $T=1~\mu$s and the amplitude of coherent states is
$|\alpha|=0.5$. With these parameters, we obtain
$\bar{F}_{\rm{NOT}}=0.9997$, indicating that the NOT gate can be
implemented almost perfectly. To show the performance of different
types of quantum gates, we plot the populations of different output
states with different input states in the implementation of
the NOT, Hadamard, and the $\pi$-phase gates in
Figs.~\ref{fig3}(a), \ref{fig3}(b), and \ref{fig3}(c), respectively. As shown, the populations of the output states are all very close to the ideal values of the theoretical results, and the leakage to unwanted levels is negligible.

{For example, we calculate the final fidelities $F_{\mathrm{H}}(T)$ of the Hadamard gate with different input states. For the input state $|\mathcal{C}_+\rangle$, we obtain
\begin{eqnarray}
&&\langle\mathcal{C}_+|U_{\mathrm{H}}(T)|\mathcal{C}_+\rangle=0.6773+0.2125i,\cr\cr
&&\langle\mathcal{C}_-|U_{\mathrm{H}}(T)|\mathcal{C}_+\rangle=0.6743+0.2036i,\nonumber
\end{eqnarray}
resulting in
\begin{eqnarray}
&&P_+(T)=|\langle\mathcal{C}_+|U_{\mathrm{H}}(T)|\mathcal{C}_+\rangle|^2=0.5039,\cr\cr
&&P_-(T)=|\langle\mathcal{C}_-|U_{\mathrm{H}}(T)|\mathcal{C}_+\rangle|^2=0.4961.\nonumber
\end{eqnarray}}

{Moreover, for the input state
$|\mathcal{C}_+\rangle$, the gate fidelity is
\begin{eqnarray}
F_{\mathrm{H}}^+(T)&=&|\langle\Psi_+(T)|\mathcal{C}_+\rangle|^2\cr\cr
&=&\frac{1}{2}|\langle\mathcal{C}_+|U_{\mathrm{H}}(T)|\mathcal{C}_+\rangle+\langle\mathcal{C}_-|U_{\mathrm{H}}(T)|\mathcal{C}_+\rangle|^2\cr\cr&=&1-4.4480\times10^{-5},\nonumber
\end{eqnarray}
where
\begin{equation}
|\Psi_+(T)\rangle=U_{\mathrm{H}}(T)|\mathcal{C}_+\rangle=\frac{1}{\sqrt{2}}(|\mathcal{C}_+\rangle+|\mathcal{C}_-\rangle).\nonumber
\end{equation}}

\begin{figure}[b]
    \centering\scalebox{0.39}{\includegraphics{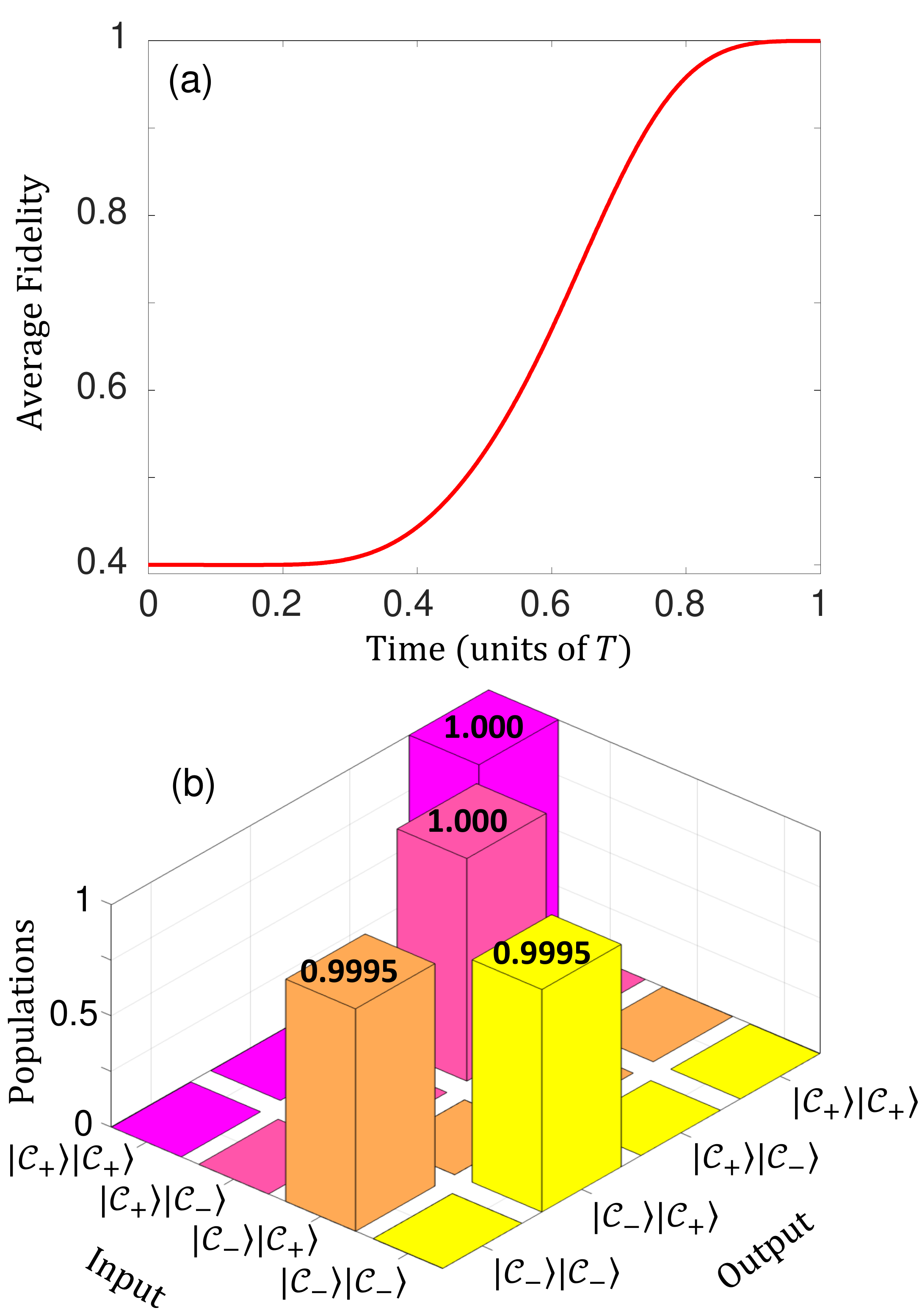}} \caption{Implementation of the CNOT gate: (a) Time variation of the average fidelity $\bar{F}_{\mathrm{CNOT}}(t)$. (b) Populations of different output states for different input states.}\label{fig4}
\end{figure}

{For the input state $|\mathcal{C}_-\rangle$, we
obtain
\begin{eqnarray}
&&\langle\mathcal{C}_+|U_{\mathrm{H}}(T)|\mathcal{C}_-\rangle=0.6743+0.2036i,\cr\cr
&&\langle\mathcal{C}_-|U_{\mathrm{H}}(T)|\mathcal{C}_-\rangle=-0.6817-0.1978i,\nonumber   \end{eqnarray}
resulting in
\begin{eqnarray}
&&P_+'(T)=|\langle\mathcal{C}_+|U_{\mathrm{H}}(T)|\mathcal{C}_-\rangle|^2=0.4961,\cr\cr
&&P_-'(T)=|\langle\mathcal{C}_-|U_{\mathrm{H}}(T)|\mathcal{C}_-\rangle|^2=0.5039.\nonumber
\end{eqnarray}
Accordingly, the gate fidelity in this case is
\begin{eqnarray}
F_{\mathrm{H}}^-(T)&=&|\langle\Psi_-(T)|\mathcal{C}_-\rangle|^2\cr\cr
&=&\frac{1}{2}|\langle\mathcal{C}_+|U_{\mathrm{H}}(T)|\mathcal{C}_-\rangle-\langle\mathcal{C}_-|U_{\mathrm{H}}(T)|\mathcal{C}_-\rangle|^2\cr\cr
&=&1-4.4752\times10^{-5},\nonumber
\end{eqnarray}
where
\begin{equation}
|\Psi_-(T)\rangle=U_{\mathrm{H}}(T)|\mathcal{C}_-\rangle=\frac{1}{\sqrt{2}}(|\mathcal{C}_+\rangle-|\mathcal{C}_-\rangle).\nonumber
\end{equation}
Therefore, the fidelities of the Hadamard gate for the two input
states $|\mathcal{C}_\pm\rangle$ are both nearly unity.}

{In addition, we also plot the average fidelities of the NOT, Hadamard and $\pi$-phase gates in Fig.~\ref{fig3}(d). The average fidelities of the three gates are $\bar{F}_{\rm{NOT}}=0.9997$, $\bar{F}_{\mathrm{H}}=0.9999$, and $\bar{F}_{\pi}=0.9998$, respectively. The results show that the three gates are implemented with very high fidelities.}

\subsubsection{Hamiltonian and evolution operator of two coupled resonators}

Now we consider a generalized scheme when two cavity modes are driven by two Kerr-nonlinear
resonators, as described by the Hamiltonian
\begin{equation}\label{e16}
H_{\mathrm{cat2}}=-K\sum_{n=1,2}a_{n}^{\dag
2}a_{n}^{2}+\epsilon_{2}(a_{n}^{\dag2}+a_{n}^2).
\end{equation}
The product coherent states
$|\pm\alpha\rangle_{1}\otimes|\pm\alpha\rangle_{2}$ of the modes $a_{1}$
and $a_{2}$ are the four degenerate eigenstates of
$H_{\mathrm{cat}2}$. Therefore, the product cat states
$\{|\mathcal{C}_\pm\rangle_{1}\otimes|\mathcal{C}_\pm\rangle_{2}\}$
span a four-dimensional computational subspace $\mathcal{S}_2$ useful for
implementing two-qubit gates.

Additionally, we consider a
control Hamiltonian
\cite{MirrahimiNJP16,GuillaudPRX9,TouzardPRL122}
\begin{align}\label{eq1-11}
H_{c2}(t)=&~\chi_{12}(t)a_{1}^{\dag} a_{1}a_{2}^{\dag}
a_{2}+a_{1}^{\dag}a_{1}[\lambda^*(t)a_{2}+\lambda(t)a_{2}^{\dag}]\cr
&+\tilde{\epsilon}^*(t)a_{2}+\tilde{\epsilon}(t)a_{2}^\dag+\sum_{n=1,2}\chi_{n}(t)a_{n}^\dag a_{n},
\end{align}
where $\chi_{12}(t)$ is the cross-Kerr parameter, $\lambda(t)$
is the strength of a resonant longitudinal interaction between the
modes $a_{1}$ and $a_{2}$; $\chi_{1/2}(t)$ are the detunings, and
$\tilde{\epsilon}(t)$ is the strength of an additional coherent
driving of the mode $a_{2}$. We assume that the parameters in $H_{\rm{c2}}(t)$
should be much smaller than the energy gap $E_{\mathrm{gap}2}$ of
the eigenstates of $H_{\rm{cat2}}$ to limit the evolution to the
subspace $\mathcal{S}_{2}$.

To realize geometric
controlled $\theta$-rotation gates, we choose the parameters
in Eq.~(\ref{eq1-11}) as follows
\begin{align}\label{eq1-12}
  \text{Re}[\lambda(t),\xi]=&~\frac{(\mathcal{N}_{+}\mathcal{N}_{-})^{\frac{3}{2}}\left(\Omega_{x}\cos\xi+e^{2|\alpha|^2}\Omega_{y}\sin\xi\right)}{8\alpha^{3}\left(\mathcal{N}_{+}^{2}-\mathcal{N}_{-}^{2}\right)},\cr
  \text{Im}[\lambda(t),\xi]=&~\text{Re}[\lambda(t),\xi-\pi/2],\cr
  \chi_{12}(t)=&~\frac{\dot{\eta}\sin^{2}\mu\, \mathcal{N}_{+}^{2}\mathcal{N}_{-}^{2}}{\left[(\mathcal{N}_{+}^{2}-\mathcal{N}_{-}^{2})^2\alpha^{4}\right]},\cr
  \chi_{1}(t)=&~-\frac{\chi_{12}(t)|\alpha|^{2}\mathcal{N}_{-}}{\mathcal{N}_{+}}, \cr
  \chi_{2}(t)=&~-\frac{\chi_{12}(t)|\alpha|^{2}{\left(\mathcal{N}_{+}^{2}+\mathcal{N}_{-}^{2}\right)}}{\left(2\mathcal{N}_{+}\mathcal{N}_{-}\right)},\cr
  \tilde{\epsilon}(t)=&~\frac{\lambda(t)\chi_{1}(t)}{\chi_{12}(t)}.
\end{align}
The evolution operator of the system reads
\begin{equation}
U_{2}(T,0)=|\mathcal{C}_+\rangle_{2}\langle\mathcal{C}_+|\otimes\mathbbm{1}_{2}+|\mathcal{C}_-\rangle_{2}\langle\mathcal{C}_-|\otimes U_{s}(T,0),
\end{equation}
where $\mathbbm{1}_{2}$ is the identity operator acting
on the cat qubit 2 and $U_{s}(T,0)$ is the single-qubit operation
acting on the cat qubit 2 defined by Eq.~(\ref{eq1-7}).

\subsection{Two-qubit entangling gates}

\begin{figure}[b]
    \centering\scalebox{0.46}{\includegraphics{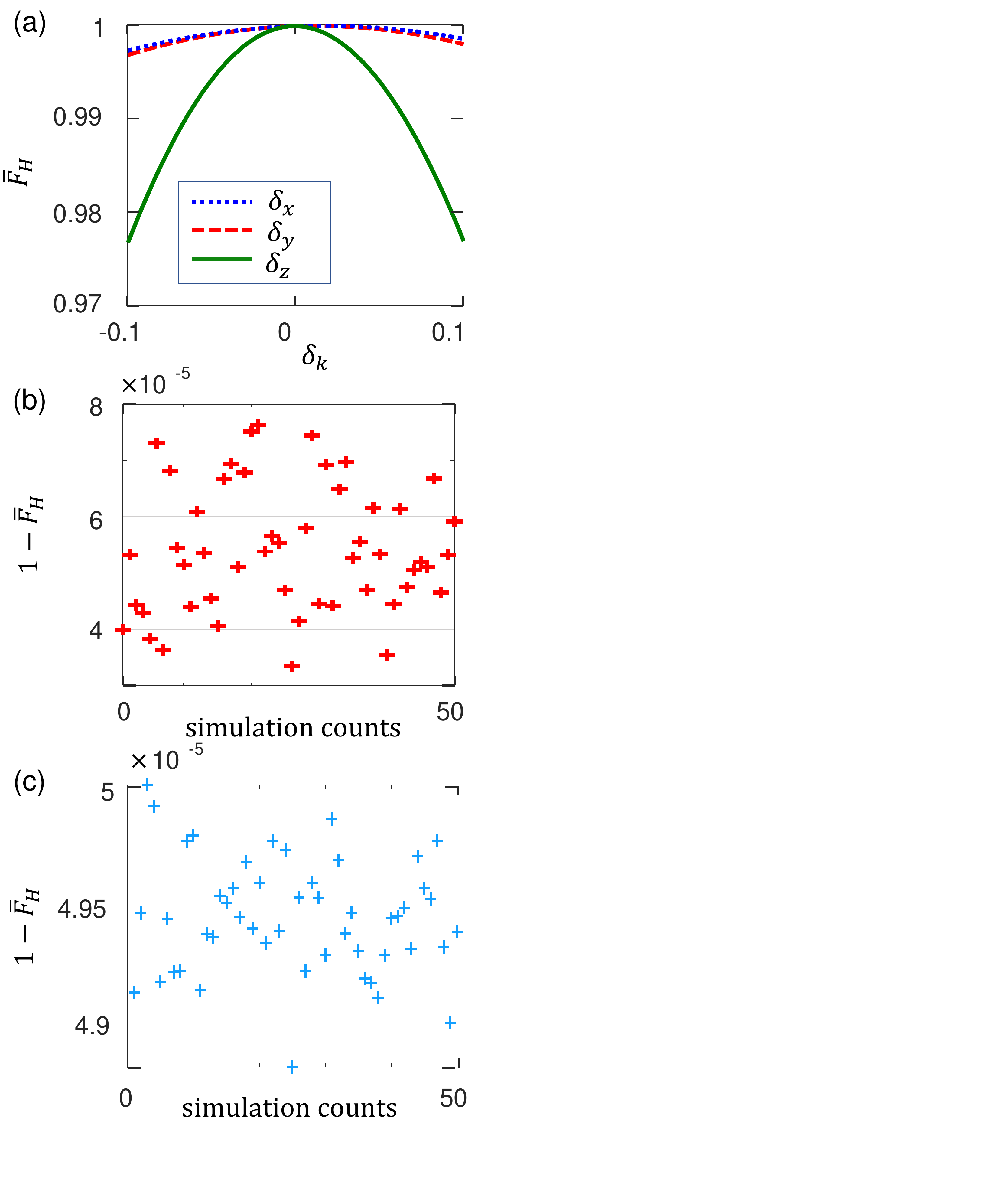}}
    \caption{(a) Final-state average fidelities $\bar{F}_{\mathrm{H}}(T)$ versus the
        systematic error coefficient $\delta_k$. (b) Final-state average infidelities, $1-\bar{F}_{\mathrm{H}}(T)$, versus simulation counts with the additive
        white Gaussian noise (signal-to-noise
ratio $R_{\rm{SN}}=10$). {(c)
Final-state average infidelities, $1-\bar{F}_{\mathrm{H}}(T)$,
versus simulation counts with $1/f$ noise (signal-to-noise ratio
$\tilde{R}_{\rm{SN}}=10$).} We set here: $K=2\pi\times12.5$~MHz and
$\alpha=0.5$.}\label{fig5}
\end{figure}

\subsubsection{Example of a two-qubit entangling gate}

As an example of the application of a two-qubit entangling gate, we
show the implementation of a modified controlled-NOT (CNOT) gate defined by the
operator
\begin{equation}
U_{\rm{CNOT}}=\mathbbm{1}_{1}\otimes|\mathcal{C}_+\rangle_{2}\langle\mathcal{C}_+|+i\sigma_{x}\otimes|\mathcal{C}_-\rangle_{2}\langle\mathcal{C}_-|,
\end{equation}
corresponding to the parameters $\eta_{0}=\pi/2$, $\mu_{0}=\pi/2$,
and $\theta=\pi/2$ for $U_{1}$. The following discussion is based on the parameters
$\alpha_1=\alpha_2=0.5$, $K=2\pi\times12.5$~MHz
($E_{\mathrm{gap}}=161$~MHz), and $T=1~\mu$s.
The average fidelity
$\bar{F}_{\mathrm{CNOT}}(t)$ of the implementation of this
CNOT gate over all possible initial states in the
computational subspace $\mathcal{S}_2$ is defined by
\cite{ZanardiPRA70,PedersenPLA367}
\begin{equation*}\label{e15}
\bar{F}_{\mathrm{CNOT}}(t)=\frac{1}{\mathcal{D}_2(\mathcal{D}_2+1)}\{\mathrm{Tr}[M_2(t)M_2^\dag(t)]
+|\mathrm{Tr}[M_2(t)]|^2\},
\end{equation*}
where
\begin{equation}
M_2(t)=\mathcal{P}_{c2}U^\dag_{\mathrm{CNOT}}U(t)\mathcal{P}_{c2},
\end{equation}
given via the projector $\mathcal{P}_{c2}$ and the dimension $\mathcal{D}_2=4$ of the computational subspace $\mathcal{S}_2$. We plot the time variation of $\bar{F}_{\mathrm{CNOT}}(t)$ in Fig.~\ref{fig4}(a), and obtain
$\bar{F}_{\mathrm{CNOT}}(T)=0.9997$. Consequently, the
modified CNOT gate can be realized almost perfectly.

Populations of different output states with different input states
in the implementation of our CNOT-like gate are plotted in
Fig.~\ref{fig4}(b). As seen, the system does
not evolve when the cavity mode 1 is in the cat state
$|\mathcal{C}_+\rangle_1$. However, if the cavity mode 1 is in the cat
state $|\mathcal{C}_-\rangle_{1}$, a nearly perfect population inversion
occurs to the cavity mode 2. The result of Fig.~\ref{fig4}(b) also
indicates that the CNOT gate is successfully
implemented with an extremely small leakage to unwanted levels.

\section{Discussions on experimental imperfections}\label{doei}

Here, we estimate the performance of the protocol in the
presence of different experimental imperfections. We consider our
implementation of a Hadamard gate as an example. First, due to an
imperfect calibration of the instruments, there may exist systematic
errors in the control parameters. The control parameter under the
influence of systematic errors can be written as
$\Omega_k^e(t)=(1+\delta_k)\Omega_k(t)$, where $\Omega_k^e(t)$ is a
faulty control parameter and $\delta_k$ is the corresponding
error coefficient.

We plot the average fidelity $\bar{F}_{\mathrm{H}}$ of the Hadamard gate
versus the systematic error coefficient $\delta_k$ in
Fig.~\ref{fig5}(a). It is seen that, when
$\delta_x\in[-10\%,10\%]$ ($\delta_y\in[-10\%,10\%]$), the average
fidelity $\bar{F}_{\mathrm{H}}$ remains higher than 0.9969 (0.9973). In addition, the
influence on $\bar{F}_{\mathrm{H}}$ caused by systematic errors of $\Omega_z(t)$, is larger
than those of $\Omega_x(t)$ and $\Omega_y(t)$, but we can still
obtain $\bar{F}_{\mathrm{H}}\geq0.9768$ when $\delta_z\in[-10\%,10\%]$. Therefore, the implementation of the Hadamard gate is robust against systematic errors.

Apart from systematic errors, due to random noise, there are also
fluctuations of parameters that may disturb the evolution of the
system. Additive white Gaussian noise (AWGN) is a good model to
investigate random processes
\cite{Dong2015,Dong2016,KYHPRA97,KYHPRA101}. Therefore, we add AWGN
to the control parameter as
\begin{equation}
\Omega_k^n(t)=\Omega_k(t)+\mathrm{AWGN}[\Omega_k(t),R_{SN}].
\end{equation}
Here, $\mathrm{AWGN}[\Omega_k(t),R_{SN}]$ is a function that generates
AWGN for the original signal $\Omega_k(t)$ with a signal-to-noise ratio $R_{\rm{SN}}$. As AWGN is generated randomly in each single simulation, we perform the numerical simulation averaged over 50 samples to estimate its average effect. Then, $1-\bar{F}_{\mathrm{H}}$
in each single simulation is plotted in Fig.~\ref{fig5}(b).
Thus, we find the values of the infidelity
$1-\bar{F}_{\mathrm{H}}\in[3\times10^{-5},8\times10^{-5}]$ averaged over the fifty
simulations. The results indicate that the implementation of the
Hadamard gate is insensitive to AWGN.

{Apart from AWGN, the $1/f$ noise in coherent drives is also a noise limiting the performance of quantum devices. When considering this noise, the control parameter becomes
\begin{equation}
\tilde{\Omega}_k^n(t)=\Omega_k(t)+\mathrm{noise}_{f^{-1}}[\Omega_k(t),\tilde{R}_{SN}],
\end{equation}
with $\mathrm{noise}_{f^{-1}}[\Omega_k(t),\tilde{R}_{SN}]$ being a function generating the
$1/f$ noise for the original signal $\Omega_k(t)$ with a signal-to-noise ratio
$\tilde{R}_{\rm{SN}}$. Here, we also perform numerical simulations averaged over 50 samples to estimate the average effect of the $1/f$ noise with $\tilde{R}_{\rm{SN}}=10$. The results are shown in Fig.~\ref{fig5}(c), where we find
$1-\bar{F}_{\mathrm{H}}\in[4.884\times10^{-5},5.004\times10^{-5}]$ averaged over the fifty simulations. Therefore, our implementation of the Hadamard gate is also insensitive to $1/f$ noise.}

\begin{figure}
    \centering\scalebox{0.5}{\includegraphics{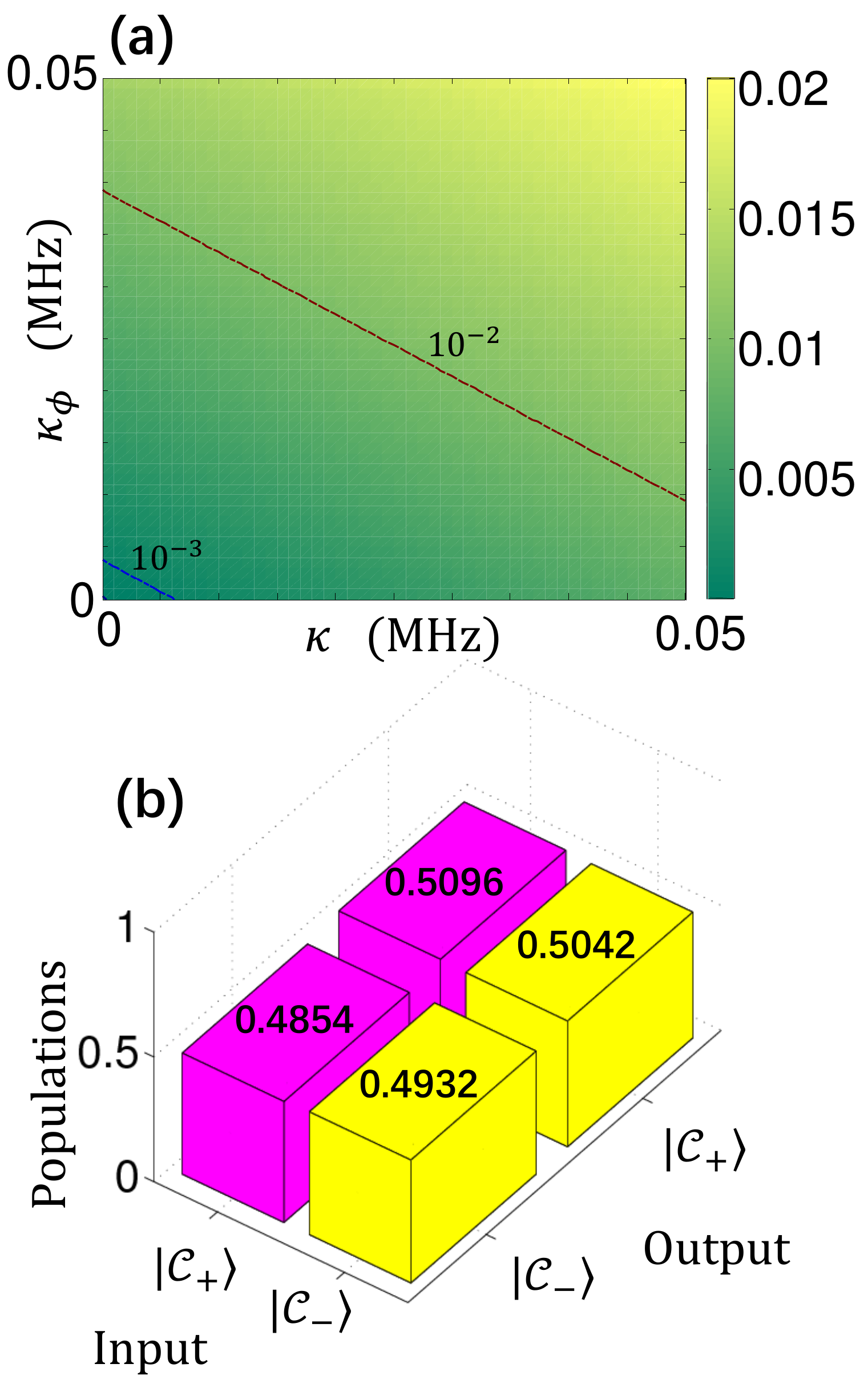}} \caption{(a)
        Infidelity $1-F_{\mathrm{H}}$ versus the single photon loss rate $\kappa$ and the
        dephasing rate $\kappa_\phi$ with parameters $K=2\pi\times12.5$~MHz and
        $\alpha=0.5$. (b) Populations of different output
        states for different input states in the implementation of the Hadamard gate
        ($\kappa=\kappa_\phi=0.05$~MHz) with parameters $K=2\pi\times12.5$~MHz and
        $\alpha=0.5$.}\label{fig6}
\end{figure}
As the system cannot be completely isolated from the environment in
experiments, the interactions between the system and the environment
may result in decoherence. We consider two types of decoherence factors, i.e., a
single-photon loss and dephasing. The evolution of the system is
described by the Lindblad master equation \cite{MirrahimiNJP16}
\begin{align}\label{dis1}
\dot{\rho}(t)=&-i[H_{\mathrm{cat}}+H_{\mathrm{add}}(t),\rho(t)]\cr &+\frac{\kappa}{2}\mathcal{L}[a]\rho(t)+\frac{\kappa_\phi}{2}\mathcal{L}[a^\dag
a]\rho(t),
\end{align}
where $\kappa$ ($\kappa_\phi$) is the single-photon-loss (dephasing)
rate and the Lindblad superoperator $\mathcal{L}$ acting on an arbitrary operator
$o$ produces $\mathcal{L}[o]\rho(t)=2o\rho(t)o^\dag-o^\dag
o\rho(t)-\rho(t)o^\dag o$. In the presence of decoherence, the
evolution is no longer unitary. For the convenience of our
discussion, we take the evolution with initial state
$|\mathcal{C}_+\rangle_{1}$ as an example and analyze the fidelities of the
Hadamard gate as
\begin{equation}
F_{\mathrm{H}}={}_{1}\langle\mathcal{C}_+|U_{\mathrm{H}}^\dag\rho(T)U_{\mathrm{H}}|\mathcal{C}_+\rangle_{1}.
\end{equation}

We plot the infidelity $1-F_{\mathrm{H}}$ versus the single photon loss rate $\kappa$ and the dephasing rate $\kappa_\phi$ in Fig.~\ref{fig6}(a) in the range [0,0.05~MHz] \cite{GiovanniniSci347}. The results show that the influence of single-photon loss is stronger than dephasing. When $\kappa,\kappa_\phi\leq0.05$~MHz, $1-F_{\mathrm{H}}(T)$
is lower than 0.0201. Therefore, the protocol is
robust against single-photon loss and dephasing. In addition,
the populations of the output states corresponding to different input states
in the implementation of the Hadamard gate with decoherence rates
$\kappa=\kappa_\phi=0.05$~MHz are plotted in Fig.~\ref{fig6}(b).
Compared with the results shown in Fig.~\ref{fig3}(a), in the
presence of decoherence, there exist more faulty populations of the
output states. This is because the single-photon loss continuously
causes quantum jumps between the cat states $|\mathcal{C}_\pm\rangle_{1}$
\cite{Purinpj3}. The total populations in the subspace $\mathcal{S}$
with the input states $|\mathcal{C}_\pm\rangle_{1}$ are both higher than
0.995, showing that the leakage to the unwanted levels outside the
subspace $\mathcal{S}$ is still very small in the presence of
decoherence.

\section{Amplification of the cat-state amplitude by squeezing the drive signal}

\begin{figure}
\centering\scalebox{0.45}{\includegraphics{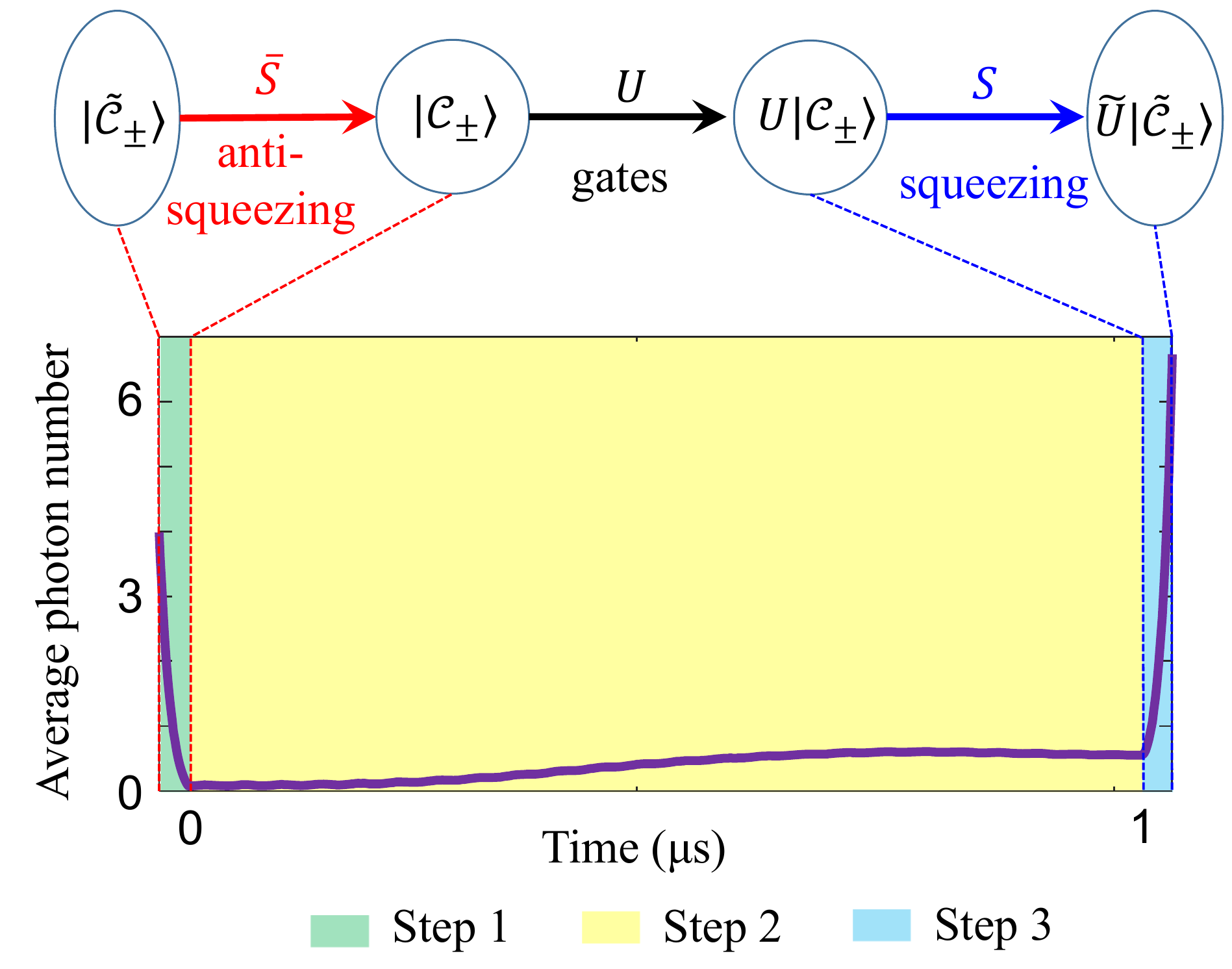}} \caption{Top panel: The
process of NGQC for the amplified cat-state qubits.
Bottom panel: Time variation of the average photon number $n_p(t)=\mathrm{Tr}[\rho(t)a^\dag a]$. Step 1 (3) is an anti-squeezing (squeezing) process for a measurable input (output) state.
Step 2 denotes evolution of the system implementing a given gate (e.g., a Hadamard gate in the bottom panel).}\label{fadd1}
\end{figure}

To increase the distinguishability of cat-state qubits, we can
introduce a method to amplify the photon numbers inspired by Refs.~\cite{SekatskiPRL103,Qin2019}.
When we consider a squeezing operator as
$S=\exp[r(a^{\dag2}-a^2)/2]$, the cat states
$\{|\mathcal{C}_\pm\rangle\}$ become the amplified cat states
$\{|\tilde{\mathcal{C}}_\pm\rangle\}$ with
$|\tilde{\mathcal{C}}_\pm\rangle=S|\mathcal{C}_\pm\rangle$ (i.e., squeezed cat states).
Here, we omit the subscript ``1'' for simplicity.
The squeezing operator $S$ can be realized by two-photon (squeezing) driving
\begin{equation}
H_s=-i\epsilon_2(a^2-a^{\dag2}),
\end{equation}
for the interaction time
$t_s=r/2\epsilon_2$, i.e., switching off the Kerr interaction and the control field [$H_{c}(t)$]. In addition, the inverse transform
$\bar{S}=S^\dag$ can be realized by squeezing the driving interaction
\begin{equation}
\bar{H}_s=-H_s=i\epsilon_2(a^2-a^{\dag2}),
\end{equation}
for the interaction time
$t_s=r/2\epsilon_2$. In this case, the total process can be
divided into three steps as shown in Fig.~\ref{fadd1}.

In step 1, we apply the transform $\bar{S}$ to the input state, which is a superposition of the squeezed cat states.
In this way, the amplified cat states
$\{|\tilde{\mathcal{C}}_\pm\rangle\}$ are transformed into
small-amplitude cat states $\{|\mathcal{C}_\pm\rangle\}$.
The step 2 is the geometric gate operation, as illustrated in Sec. III.
In step 3, we apply the squeezing
transform $S$ to enhance the photon number of the output state, so that
the output state can be experimentally detected. The total
operator acting on the amplified cat-state qubit is
$\tilde{U}=SUS^\dag$, which has the same matrix elements for
a small-amplitude cat-state qubit, i.e.,
$\langle\tilde{\mathcal{C}}_\jmath|\tilde{U}|\tilde{\mathcal{C}}_{\jmath'}\rangle=\langle
\mathcal{C}_\jmath|U|\mathcal{C}_{\jmath'}\rangle$
($\jmath,\jmath'=\pm$). Here, we assume $r=1.2$ and $\alpha=0.5$, as an
example to show the implementation of NGQC for the amplified-cat-state
qubits.

We plot the time variation of the average photon
number $n_p(t)=\mathrm{Tr}[\rho(t)a^\dag a]$ in the bottom panel of Fig.~\ref{fadd1}. As shown, the average photon number decreases during the first step, corresponding to the anti-squeezing process $|\tilde{\mathcal{C}}_+\rangle\rightarrow|\mathcal{C}_+\rangle$. Then, by implementing the Hadamard gate, the cat
state $|\mathcal{C}_+\rangle_{1}$ is transformed into
$(|\mathcal{C}_+\rangle+|\mathcal{C}_-\rangle)/\sqrt{2}$. Finally, in step 3, we amplify the output state by squeezing, i.e., 
\begin{equation}
\frac{1}{\sqrt{2}}\left(|\mathcal{C}_+\rangle+|\mathcal{C}_-\rangle\right)\rightarrow\frac{1}{\sqrt{2}}\left(|\tilde{\mathcal{C}}_+\rangle+|\tilde{\mathcal{C}}_-\rangle\right).\nonumber    
\end{equation}
The final average photon number is 6.732. The anti-squeezing and squeezing processes are fast (see Fig.~\ref{fadd1}), so that decoherence in these two processes affects weakly the target state. By considering the experimentally feasible parameters:
$\epsilon_2=K\alpha^2=2\pi\times3.125$~MHz,
$t_s=r/\epsilon_2=30.56$~ns, and $\kappa=\kappa_\phi=0.05$~MHz, we achieve the
fidelity $F_\mathrm{H}=0.9513$ of the output state $(|\tilde{\mathcal{C}}_+\rangle+|\tilde{\mathcal{C}}_-\rangle)/\sqrt{2}$ for the Hadamard gate with the initial state
$|\tilde{\mathcal{C}}_+\rangle$ for the amplified cat-state qubit.

\section{Possible implementation using superconducting quantum circuits}

\subsection{Single-qubit geometric quantum gates}

\begin{figure}
    \centering\scalebox{0.5}{\includegraphics{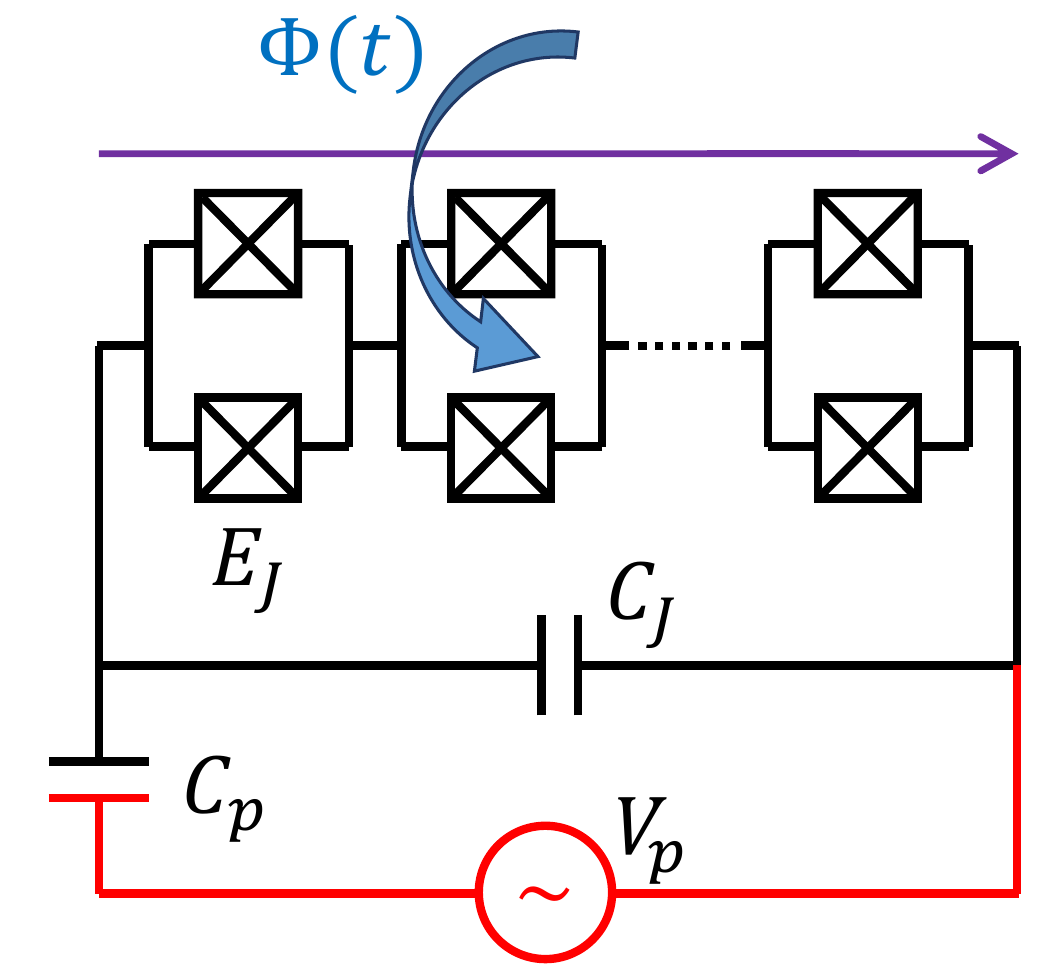}}
    \caption{uperconducting quantum circuit for implementing Hamiltonian in Eq.~(\ref{e2.5}). The circuit consists of a SQUID
        array (black), a shunting capacitor (black),
        a flux bias line (purple), and an ac gate voltage (red). Here, $\Phi(t)$ is the externally applied magnetic flux, $C_{J}$ is the capacitor shunting the SQUID array, $E_{J}$ is the Josephson energy of a single SQUID, $V_{p}$ is the amplitude of the ac gate voltage, and $C_p$ is the gate capacitor.}
    \label{model}
\end{figure}

As shown in Fig.~\ref{model}, we consider an array-type
resonator composed of $N$ superconducting quantum interference devices (SQUIDs) \cite{Yaakobi2013,You2011,GU20171,Kockum2019,WangPRX2019,Kjaergaard2020,arXiv200905723}.
An ac gate voltage $\tilde{V}_{p}=V_{p}\cos(\omega_{p}t+\varphi_{p})$ (with amplitude $V_{p}$, frequency $\omega_{p}$, and phase $\varphi_{p}$) is applied to induce
linear transitions between eigenlevels. The Hamiltonian of this setup reads \cite{GU20171,WangPRX2019,arXiv200905723}
\begin{eqnarray}
  H_{0}=4E_{C}{\hat{n}}^2-NE_{J}[\Phi(t)]\cos\left(\frac{\hat{\phi}}{N}\right)-\frac{E_{C}C_{p}\tilde{V}_p}{e}\hat{n},\ \
\end{eqnarray}
where $\hat{n}$ is the number of Cooper pairs and $\hat{\phi}$ is
the overall phase across the junction array. Here,
$E_{C}$ is the resonator charging energy, $E_{J}$ is the Josephson energy of a single SQUID,
and $N$ is the number of SQUIDs in the array. The Josephson energy is periodically modulated
(with frequency $\omega_{2p}$ and phase $\varphi_{2p}$) by the external
magnetic flux $\Phi(t)$, leading to
\begin{equation}
E_{J}[\Phi(t)]=E_{J}+\tilde{E}_{J}\cos(\omega_{2p}t+\varphi_{2p}).
\end{equation}

After the Taylor expansion of $\cos(\hat{\phi}/N)$ to
fourth order, we obtain
\begin{align}\label{mod1}
  H_{0}\approx &~ 4E_{C}\hat{n}^2-NE_{J}(1-\hat{X}+\hat{X}^2/6)\cr
  &- N\tilde{E}_{J}(1-\hat{X})\cos(\omega_{2p}t+\varphi_{2p})\cr
  &-\frac{E_{C}C_{p}\tilde{V}_{p}}{e}\hat{n},
\end{align}
where $\hat{X}=(\hat{\phi}/N)^2/2$. The quadratic time-independent part of the Hamiltonian
can be diagonalized by defining
\begin{align}
  \hat{n}=-i n_{0}(a-a^{\dag}), \ \ \hat{\phi}=\phi_{0}(a+a^{\dag}),
\end{align}
where $n_{0}=[E_{J}/(32NE_{C})]^{1/4}$ and ${\phi_{0}}=1/2n_0$ are the
zero-point fluctuations.

By dropping the constant terms, the Hamiltonian $H_{0}$ becomes
\begin{align}
  H_{0}=&~\omega_{c}a^{\dag}a-\frac{E_{C}}{12N^{2}}\left(a+a^{\dag}\right)^{4}\cr
        &+\frac{\tilde{E}_{J}\omega_{c}}{4E_{J}}\left(a+a^{\dag}\right)^2 \cos(\omega_{2p}t+\varphi_{2p}) \cr
        &+i\frac{E_{C}C_{p}{V}_{p}}{e}(a-a^{\dag})\cos(\omega_{p}t+\varphi_{p}),
\end{align}
where $\omega_{c}=\sqrt{8E_{C}E_{J}/N}$.
For simplicity, we assume $\omega_{2p}=2\omega_{p}$ and $\varphi_{2p}=-2\xi$. Then, moving into a rotating frame at frequency $\omega_{p}$
and neglecting all of the fast oscillating terms,
the approximate Hamiltonian under the rotating wave approximation (RWA) can be written as
\begin{align}
H_{0}=&-Ka^{\dag
2}a^{2}+\epsilon_2\left(e^{2i\xi}a^{\dag2}+e^{-2i\xi}a^2\right)\cr
      &+\chi(t) a^{\dag}a+\epsilon(t)a^{\dag}+\epsilon^{*}(t)a,
\end{align}
where $K=E_{C}/2N^{2}$, $\epsilon_{2}=\omega_{c}\tilde{E}_{J}/8E_{J}$, $\chi(t)=\omega_{c}-\omega_{p}$, and
\begin{equation}
\epsilon(t)=-i\frac{E_{C}C_{p}{V}_{p}}{2e}\exp(-i\varphi_{p}).
\end{equation}
Then, by defining $a_{1}\equiv a$, we recover
the total Hamiltonian $H_{\rm{tot}}(t)$ given in Eq.~(\ref{e2.5}) for the single-qubit case. Note that, as shown in~\cite{Bartkowiak2014}, the Kerr
nonlinearity (which is a rescaled third-order susceptibility of a
nonlinear medium) can be exponentially enhanced by applying
quadrature squeezing, which interaction with a medium is
proportional to its second-order susceptibility. This means that
one can exponentially amplify higher-order nonlinearities by
applying lower-order nonlinear effects. Other methods of applying
quadrature squeezing to increase nonlinear interactions are
described in, e.g.,~\cite{Qin2018,Leroux2018,Qin2020}.

\subsection{Errors in the sperconducting circuit}

{According to the above analysis, the control drives depend on the parameters $E_C$, $E_J$, and $V_p$. Assuming that the parameters with errors respectively become
\begin{eqnarray}\nonumber
E_C\rightarrow E_C+\delta\!E_C,\
E_J\rightarrow E_J+\delta\!E_J,\
V_p\rightarrow V_p+\delta\!V_p.
\end{eqnarray}
The errors in the control parameters can be approximately calculated by
\begin{eqnarray}\nonumber
&&\delta\!\omega_c\simeq\frac{\omega_c}{2}(\frac{\delta\!E_J}{E_J}+\frac{\delta\!E_C}{E_C}),\ \delta\!K=\frac{K\delta\!E_C}{E_C},\cr\cr &&\delta\!\epsilon_2\simeq\frac{\epsilon_2}{2}\left(\frac{\delta\!E_C}{E_C}-\frac{\delta\!E_J}{E_J}\right),\cr\cr &&\delta\!\epsilon(t)\simeq\left(\frac{\delta\!E_C}{E_C}+\frac{\delta\!V_p}{V_p}\right)\epsilon(t).
\end{eqnarray}
Therefore, when considering the errors in the control parameters, the Hamiltonian becomes
\begin{align}\label{nod1}
H_{0}'=&H_0+\delta\!H_{\rm{cat}}+\delta\!H_c,\cr
\delta\!H_{\rm{cat}}=&-\delta\!Ka^{\dag
2}a^{2}+\delta\!\epsilon_2\left(e^{2i\xi}a^{\dag2}+e^{-2i\xi}a^2\right),\cr
\delta\!H_c=&\delta\!\omega_ca^{\dag}a+\delta\!\epsilon(t)a^{\dag}+\delta\!\epsilon^{*}(t)a,
\end{align}
where $\delta\!H_{\rm{cat}}$ induces an error on the amplitude of the coherence states $|\!\pm\alpha\rangle$ as
\begin{equation}
\delta\!\alpha\simeq-\frac{\alpha}{4}\left(\frac{2\delta\!E_C}{E_C}-\frac{\delta\!E_J}{E_J}\right),\nonumber
\end{equation}
and $\delta\!H_c$ influences the control drives designed by invariant-based
reverse engineering. }

{For $|\delta\!E_C/E_C|,|\delta\!E_J/E_J|\leq0.1$, and $\alpha=0.5$, the absolute error of the coherent-state amplitude approximately satisfies $|\delta\!\alpha|\leq0.0375$. Because of
\begin{equation}
1-|\langle\tilde{\alpha}_+|\alpha\rangle|^2\simeq1-|\langle\tilde{\alpha}_-|\alpha\rangle|^2\simeq0.001,\nonumber
\end{equation}
with $\tilde{\alpha}_\pm=0.5\pm0.0375$, the cat qubit can be near perfectly stabilized in the preset states by a small fluctuation of the amplitude $\alpha$.}

{Moreover, according to Eq.~(\ref{eq1-4}) and $\delta\!H_c$ in Eq.~(\ref{nod1}), the errors in $\Omega_x(t)$, $\Omega_y(t)$, and $\Omega_z(t)$ can be approximately described by the linear superpositions of $\delta\!E_J$, $\delta\!E_C$, and $\delta\!V_p$. Therefore, when $E_J$, $E_C$, and $V_p$ are subjected to AWGN or $1/f$ noise, the wave shapes of $\Omega_x(t)$, $\Omega_y(t)$, and $\Omega_z(t)$ are also mixed with the same type of noise. According to the analysis in Sec.~\ref{doei}, the fidelities of the gates are insensitive to these types of noise in $\Omega_x(t)$, $\Omega_y(t)$, and $\Omega_z(t)$. Consequently, our proposal is robust against AWGN and $1/f$ noise in $E_J$ and $E_C$ of the considered superconducting circuits.}

\subsection{Kitten states}

We also note that the Kerr nonlinearity enables the generation of not only
conventional (i.e., two-component) Schr\"odinger cat states, i.e.,
superpositions of two macroscopically distinct states, but also
the generation of superpositions of a larger number of
macroscopically distinct states. The states are referred to as
Schr\"odinger kitten states or multi-component cat-like states, as
predicted in~\cite{Miranowicz1990} and experimentally generated
via the Kerr interaction in superconducting quantum circuits
in~\cite{Kirchmair2013}. These kitten states, which are examples
of Gauss sums, have been used in an unconventional algorithm for
number factorization, i.e., to distinguish between factors and
nonfactors. Implementations of the Gauss-sums algorithm include
NMR spectroscopy~\cite{Mehring2007,Mahesh2007} and Ramsey
spectroscopy using cold atoms~\cite{Gilowski2008}.

\subsection{Two-qubit geometric entangling gates}

{To realize two-qubit geometric gates, we consider two superconducting circuits with the same structure shown in Fig.~\ref{model}. The two circuits are coupled with each other through a Josephson-junction coupler with Josephson energy $\bar{E}_J$ \cite{Vrajitoarea19,TouzardPRL122}. The coupler provides a coupling between the two circuits as
\begin{equation}
\mathcal{U}_J=-\bar{E}_J\cos\left[\hat{\phi}_2-\hat{\phi}_1+2\pi\frac{\bar{\Phi}(t)}{\Phi_0}\right],
\end{equation}
with $\hat{\phi}_m=\phi_{0m}(a_m+a^{\dag}_m)$ ($m=1,2$). Here, $\bar{\Phi}(t)$ is the external flux applied to the loop of the coupler, and $\Phi_0$ is the flux quantum. By modulating $\bar{\Phi}(t)$ at the frequency of the second circuit, the required cross-Kerr nonlinearity $\chi_{ab}(t)a^\dag_1a_1a^\dag_2a_2$ and the longitudinal interaction $a^\dag_1a_1[\lambda^*_1(t)a_2+\lambda_1(t)a^\dag_2]$ can be realized with the Taylor expansion of $\cos[\hat{\phi}_2-\hat{\phi}_1+2\pi\bar{\Phi}(t)/\Phi_0]$ to
fourth order \cite{TouzardPRL122}. The coupling between two circuits changes the detuning and strength of the two-photon drive of each circuit. However, to diagonalize the quadratic time-independent part of the Hamiltonian, the coefficients of $\hat{n}_m^2$ and $\hat{\phi}_m^2$ should be set equal, where $\hat{n}_m=-in_{0m}(a_m-a^{\dag}_m)$. Consequently, one can derive that the strength of the self-Kerr nonlinearity for each circuit is still expressed by $K=E_{C}/2N^{2}$.}

\subsection{Comparison of gate times}

{We now compare the gate time of our proposal with previous experiments. Relevant information is also listed in Table~\ref{tab2}.}

{The gate time in our proposal is comparable with that in previous experiments using superconducting systems. For example, the gate time of a controlled-$\pi$-phase gate in Ref.~\cite{Xu2020} using superconducting qubits is about 113~ns, with gate fidelities $\sim0.99$. In Ref.~\cite{Xue2017}, the gate time for implementing the Hadamard gate and the NOT gate with fidelities $\sim0.99$ (by using a transmon qubit coupled to a transmission-line resonator) is $\sim62.5$~ns. In our proposal, to implement the NOT and Hadamard gates with fidelity higher than 0.99, the gate time can be selected as $T=210$~ns. To implement the CNOT gate with fidelity higher than 0.99, the gate time in the present scheme is about 220~ns.}

{Moreover, compared with some previous methods using dipole interactions between neutral atoms, the gate time for two-qubit gates in the present approach is generally shorter. For example, in Ref.~\cite{KYHPRA97}, the gate time to realize the controlled-$\pi$-phase gate with about $10^{-4}$ infidelity is 376.16~$\mu$s using the available dipole interaction strengths $V=20$~MHz. In the scheme of Ref.~\cite{KYHPRA102}, the gate time to implement the CNOT gate for two neutral atoms with infidelity about $10^{-3}$ is 85.94~$\mu$s, using the reported dipole interaction strengths $V=2\pi\times50$~MHz.}

{In the present protocol, the gate times for implementing the controlled-$\pi$-phase gate and the CNOT gate, with infidelities about $10^{-4}$ to $10^{-3}$, can be both less than 1~$\mu$s with the available Kerr nonlinearity $K=2\pi\times12.5$~MHz. The longer gate times of the controlled-$\pi$-phase gate and the CNOT gate in the schemes of Refs.~\cite{KYHPRA97,KYHPRA102} is because these protocols should work in the Rydberg blockade regime, which limits the strength of the control fields.}

{Recently, some modified nonadiabatic geometric quantum
computation methods \cite{Guo2020,Wang2021,Sun2021,Wu2021}, based on the dipole interactions of neutral atoms, have been proposed, which work in different regimes and relax the limitation of strong amplitudes of control fields. The gate times of the controlled-$\pi$-phase gate and the CNOT gate in these schemes can be improved to about 1--10~$\mu$s. In the present protocol, as the Kerr nonlinearity can provide much bounder energy gap between eigenvectors of $H_{\mathrm{cat}}$, the control fields with stronger amplitudes can be adopted.}

{In addition, as reported in the scheme of Ref.~\cite{Huang2018}, to implement multi-atom geometric quantum gates in cavity quantum electrodynamics system with the amplitude of laser pulses about 50-200~MHz with fidelity over 0.999, the gate time is about 0.5--3~$\mu$s, which is also close to the gate time in the present approach.}

{Compared with the present method, the gate time of the nonadiabatic geometric quantum computation schemes \cite{Liang2014,Zhao2019} in trapped-ion systems is slower. Because these schemes work in the Lamb-Dicke limit, the amplitudes of driving pulses should be much less than the frequency (1--10~MHz) of the vibration modes of trapped ions, and the gate time is about 1--2~ms.}
\begin{table*}
\caption{{Comparison of the gate times between previously reported schemes and ours}}\label{tab2}
{\begin{tabular}{cccccc}
\hline\hline \ \ \ Year \ \ \ \ &\ \ \ Reference \ \ \ \ &\ \ \ System\ \ \ &\ \ \ Gate\ \ \ &\ \ \ Gate time\ \ \ &\ \ \ Fidelity\ \ \ \\
\hline \ \ \ 2014\ \ \ &\ \ \ Ref.~\cite{Liang2014}\ \ \ &\ \ \ Trapped ions\ \ \ &\ \ \ Hadamard, Phase\ \ \ &\ \ \ $\sim$1--2~ms\ \ \ &\ \ \ $\gtrsim0.99$\ \ \ \\
\ \ \ 2017\ \ \ &\ \ \ Ref.~\cite{Xue2017}\ \ \ &\ \ \ Superconducting circuits\ \ \ &\ \ \ NOT, Hadamard\ \ \ &\ \ \ $\sim62.5$~ns\ \ \ &\ \ \ $\gtrsim0.99$\ \ \ \\
\ \ \ 2018\ \ \ &\ \ \ Ref.~\cite{KYHPRA97}\ \ \ \ &\ \ \ Rydberg atoms\ \ \ &\ \ \ Controlled-$\pi$-phase\ \ \ &\ \ \ $\sim376.16$~$\mu$s\ \ \ &\ \ \ $\gtrsim0.999$\ \ \ \\
\ \ \ 2018\ \ \ &\ \ \ Ref.~\cite{Huang2018}\ \ \ &\ \ \ Cavity QED\ \ \ &\ \ \ Controlled-$\pi$-phase, Toffoli\ \ \ &\ \ \ $\sim$0.5--3~$\mu$s\ \ \ &\ \ \ $\gtrsim0.999$\ \ \ \\
\ \ \ 2019\ \ \ &\ \ \ Ref.~\cite{Zhao2019}\ \ \ &\ \ \ Trapped ions\ \ \ &\ \ \ CNOT\ \ \ &\ \ \ $\sim$1--2~ms\ \ \ &\ \ \ $\gtrsim0.99$\ \ \ \\
\ \ \ 2020\ \ \ &\ \ \ Ref.~\cite{Xu2020}\ \ \ &\ \ \ Superconducting circuits\ \ \ &\ \ \ Controlled-$\pi$-phase\ \ \ &\ \ \ $\sim113$~ns\ \ \ &\ \ \ $\gtrsim0.99$\ \ \ \\
\ \ \ 2020\ \ \ &\ \ \ Ref.~\cite{KYHPRA102}\ \ \ \ &\ \ \ Rydberg atoms \ \ \ &\ \ \ CNOT\ \ \ &\ \ \ $\sim85.94$~$\mu$s\ \ \ &\ \ \ $\gtrsim0.999$\ \ \ \\
\ \ \ 2020\ \ \ &\ \ \ Ref.~\cite{Guo2020}\ \ \ &\ \ \ Rydberg atoms\ \ \ &\ \ \ Controlled-$\pi$-phase, CNOT\ \ \ &\ \ \ $\sim8$~$\mu$s\ \ \ &\ \ \ $\gtrsim0.99$\ \ \ \\
\ \ \ 2021\ \ \ &\ \ \ Ref.~\cite{Wang2021}\ \ \ &\ \ \ Rydberg atoms\ \ \ &\ \ \ SWAP\ \ \ &\ \ \ $\sim7.7$~$\mu$s\ \ \ &\ \ \ $\gtrsim0.99$\ \ \ \\
\ \ \ 2021\ \ \ &\ \ \ Ref.~\cite{Wu2021}\ \ \ &\ \ \ Rydberg atoms\ \ \ &\ \ \ CNOT\ \ \ &\ \ \ $\sim1$~$\mu$s\ \ \ &\ \ \ $\gtrsim0.99$\ \ \ \\
\multicolumn{2}{c}{\ \ \ Our proposal\ \ \ }&\ \ \ Superconducting circuits\ \ \ &\ \ \ NOT, Phase, Hadamard, CNOT\ \ \ &\ \ \ $\sim$210--220~ns\ \ \ &\ \ \ $\gtrsim0.99$ \ \ \ \\
\hline\hline
\end{tabular}}
\end{table*}

\section{Conclusion}

In conclusion, we have proposed a method to realize nonadiabatic geometric quantum
computation using
cat qubits with invariant-based reverse
engineering. The evolution of the cavity mode is restricted to a
subspace spanned by a pair of Schr{\"o}dinger cat states assisted by a Kerr
nonlinearity and a two-photon squeezing drive, so that one can generate photonic cat qubits. We add a coherent field to linearly drive the
cavity mode, inducing oscillations between dressed cat states.
When designing the control fields by invariant-based reverse engineering, the system can evolve quasiperiodically and acquire only pure geometric phases.
Thus, one can realize nonadiabatic geometric quantum
computation with a cat qubit. By amplifying
the amplitudes of different cat states, the input and output states can be easily detected in experiments.

Two-qubit quantum gates for cat qubits are also considered with couplings between two cavity modes. As we have shown, the controlled two-qubit geometric quantum entangling gates can also be implemented with high fidelities. The influence of systematic errors, AWGN, $1/f$ noise, and decoherence (including photon loss and dephasing), was studied here using numerical simulations. The results indicate that our approach is robust against these errors. Therefore, our protocol can provide efficient high-fidelity quantum gates for nonadiabatic geometric quantum
computation in bosonic systems.

\begin{acknowledgments}
We thank Jiang~Zhang for helpful discussions.
Y.-H.C. is supported by the Japan Society for the Promotion of Science (JSPS) KAKENHI Grant No.~JP19F19028.
X.W. is supported by the China
Postdoctoral Science Foundation No.~2018M631136,
and the Natural Science Foundation of China under
Grant No.~11804270. S.-B.Z is supported by
the National Natural Science Foundation of China under Grants No.
11874114.
Y.X. is supported by the National Natural Science
Foundation of China under Grant No. 11575045, the Natural Science Funds for
Distinguished Young Scholar of Fujian Province under
Grant 2020J06011 and Project from Fuzhou University
under Grant JG202001-2.
A.M. is supported by the Polish National Science Centre (NCN)
under the Maestro Grant No.~DEC-2019/34/A/ST2/00081.
F.N. is supported in part by:
Nippon Telegraph and Telephone Corporation (NTT) Research,
the Japan Science and Technology Agency (JST) [via
the Quantum Leap Flagship Program (Q-LEAP),
and the Moonshot R\&D Grant No.~JPMJMS2061,
the Japan Society for the Promotion of Science (JSPS)
[via the Grants-in-Aid for Scientific Research (KAKENHI) Grant No. JP20H00134,
the Army Research Office (ARO) (Grant No.~W911NF-18-1-0358),
the Asian Office of Aerospace Research and Development (AOARD) (via Grant No.~FA2386-20-1-4069), and
the Foundational Questions Institute Fund (FQXi) via Grant No.~FQXi-IAF19-06.
\end{acknowledgments}

\appendix

\section{Derivation of a dynamic invariant and the choice of parameters for eliminating dynamical phases}

Because the Hamiltonian
$H_c(t)=\vec{\Omega}(t)\cdot\vec{\sigma}$ possesses SU(2) dynamic
structure, there is a dynamic invariant $I(t)$ in form of
$I(t)=\vec{\zeta}(t)\cdot\vec{\sigma}$
\cite{TorronteguiPRA89,KYHPRA101}. The commutative relation of
$H_c(t)$ and $I(t)$ can be calculated as
\begin{eqnarray}\label{apa}
[H_c(t),I(t)]
&=&~[\vec{\Omega}(t)\cdot\vec{\sigma}][\vec{\zeta}(t)\cdot\vec{\sigma}]-[\vec{\zeta}(t)\cdot\vec{\sigma}][\vec{\Omega}(t)\cdot\vec{\sigma}]\cr\cr
&=&~\{\vec{\sigma}\times[\vec{\Omega}(t)\times\vec{\zeta}(t)]\}\cdot\vec{\sigma}\cr\cr
&=&~(\vec{\sigma}\times\vec{\sigma})\cdot[\vec{\Omega}(t)\times\vec{\zeta}(t)]\cr\cr
&=&~2i[\vec{\Omega}(t)\times\vec{\zeta}(t)]\cdot\vec{\sigma}.
\end{eqnarray}
Substituting Eq.~(\ref{apa}) into Eq.~(\ref{lr1}), we obtain
$\dot{\vec{\zeta}}(t)=2[\vec{\Omega}(t)\times\vec{\zeta}(t)]$.
Moreover,
\begin{align}\label{apa2}
\frac{1}{2}\frac{d}{dt}|\vec{\zeta}(t)|^2=&~\vec{\zeta}(t)\cdot\dot{\vec{\zeta}}(t)
=2\vec{\zeta}(t)\cdot[\vec{\Omega}(t)\times\vec{\zeta}(t)]\cr
=&~2\vec{\Omega}(t)\cdot[\vec{\zeta}(t)\times\vec{\zeta}(t)]=0,
\end{align}
implies that $|\vec{\zeta}(t)|$ should be constant.

When $|\vec{\zeta}(t)|=1$, one can parametrize $\vec{\zeta}(t)$ as
$(\sin{\eta}\sin{\mu},\cos{\eta}\sin{\mu},\cos{\mu})$, and derive
the eigenvectors $|\phi_\pm(t)\rangle$ of the invariant $I(t)$ as
given in Eq.~(\ref{eq1-5}). The time derivatives of the dynamical
phases $\vartheta_\pm(t)$ and the geometric phases $\Theta_\pm(t)$
acquired by $|\phi_\pm(t)\rangle$ are
\begin{align}\label{apa3}
&\dot{\vartheta}_\pm(t)=\mp\left[\Omega_z(t)+\frac{1}{2}\dot{\eta}\sin^2\mu\right]\sec\mu,\cr
&\dot{\Theta}_\pm(t)=\pm\dot{\eta}\sin^2\left(\frac{\mu}{2}\right).
\end{align}
To eliminate the dynamical phases, we choose
\begin{equation}\label{apa4}
\Omega_z(t)=-\frac{1}{2}\dot{\eta}(t)\sin^2[\mu(t)].
\end{equation}
In addition, by reversely solving
$\dot{\vec{\zeta}}(t)=2[\vec{\Omega}(t)\times\vec{\zeta}(t)]$, we
obtain
\begin{align}\label{apa5}
\Omega_x(t)=&~\frac{1}{2}\vec{e}_z\cdot[\vec{\nu}_1(t)\times\vec{\nu}_2(t)],\cr
\Omega_y(t)=&~\frac{1}{2}\vec{\nu}_1(t)\cdot\vec{\nu}_2(t),\cr
\vec{\nu}_1(t)=&~[2\Omega_z(t)+\dot{\eta}(t)]\tan[\mu(t)]\vec{e}_x+\dot{\mu}(t)\vec{e}_y,\cr
\vec{\nu}_2(t)=&~\cos[\eta(t)]\vec{e}_x+\sin[\eta(t)]\vec{e}_y.
\end{align}
Combining Eqs.~(\ref{apa4}) and (\ref{apa5}), we derive
Eq.~(\ref{eq1-4}) as shown in Sec.~\ref{III}.


\bibliography{references}

\end{document}